\documentclass[aps,pra,twocolumn,groupedaddress]{revtex4}
\usepackage{graphicx}
\usepackage{float}
\usepackage{dcolumn}
\usepackage{amsmath}
\usepackage{amssymb}
\input{epsf}

\newcommand{\be}{\begin{equation}}
\newcommand{\ee}{\end{equation}}
\newcommand{\ba}{\begin{array}}
\newcommand{\ea}{\end{array}}

\newcommand{\calH}{{\cal H }}

\newcommand{\calB}{{\cal B }}
\newcommand{\calF}{{\cal F }}

\newcommand{\ZZ}{\mathbb{Z}}
\newcommand{\CC}{\mathbb{C}}

\newcommand{\la}{\langle}
\newcommand{\ra}{\rangle}
\newcommand{\cop}{\hat{c}}

\newcommand{\nn}{\nonumber}

\newcommand{\trace}{\mathop{\mathrm{Tr}}\nolimits}

\newcommand{\sm}{\sigma}
\newcommand{\ep}{\epsilon}

\renewcommand{\ss}{{\bf s}}
\newcommand{\rr}{{\bf r}}
\newcommand{\uu}{{\bf u}}
\renewcommand{\tt}{{\bf t}}

\newcommand{\phase}{e^{i\frac{\pi}4}}
\newcommand{\ssigma}{{\bf \sigma}}
\newcommand{\ppsi}{{\bf \psi}}
\newcommand{\one}{{\bf 1}}
\newcommand{\zero}{{\bf 0}}

\newtheorem{lemma}{Lemma}

\begin{document}

\title{Universal Quantum Computation with the $\nu=5/2$ Fractional
Quantum Hall  State}

\author{Sergey Bravyi\footnote{This research was carried out while
the author was at the Institute for Quantum Information, Caltech.}}

\affiliation{\mbox{IBM Watson Research Center}\\
\mbox{Yorktown Heights, NY 10598 USA}}


\date{\today}

\begin{abstract}
We consider topological quantum computation (TQC) with
a particular class of anyons that are believed to exist in the
Fractional Quantum Hall Effect state
at Landau level filling fraction $\nu=5/2$.
Since the braid group representation
describing statistics of these anyons
is not computationally universal, one cannot directly apply
the standard TQC technique.
We propose to use very noisy non-topological operations such as
direct short-range interaction between anyons to simulate a universal
set of gates.
Assuming that all TQC operations are implemented perfectly,
we prove that
the threshold error rate for non-topological operations
is above $14\%$. The total number of non-topological computational
elements that one needs to simulate a quantum circuit with $L$ gates
scales as $L(\log L)^3$.
\end{abstract}

\maketitle

\section{Introduction}
\label{sec:intro}

One of the most important results in the theory of fault-tolerant
quantum computation is the threshold theorem. It asserts that
ideal quantum circuits can be efficiently simulated by noisy circuits
if an error rate of individual gates is smaller than
a certain constant threshold value $\delta$,
see~\cite{Sho96a,AB96a,Kit97b,AGP05a}.
Estimates of $\delta$ vary
from $10^{-7}-10^{-4}$ for a local
architecture~\cite{Roy04,STD04a}
to $10^{-5}-10^{-2}$ for non-local gates~\cite{AGP05a,Kni05n}.
With the present technology these rates are hardly achievable by any
real device. Moreover, for practical computations it is desirable to
have an error rate much smaller than $\delta$, otherwise
one may need too many concatenation levels and
the simulation overhead may become too large.

These challenges can be overcomed (at least partially)
in the {\it Topological Quantum Computation} (TQC) scheme
developed by Kitaev, Freedman, and
Preskill~\cite{Kit97a,FKLW01a,DKLP02a}.
It makes use of the fact that elementary excitations of
some two-dimensional many-body quantum systems are {\it anyons}
--- spacially localized quasiparticles with unusual exchange
statistics described by non-trivial representations of the
braid group. For the purposes of TQC one needs {\it non-Abelian
anyons} (corresponding to multi-dimensional braid group
representations). A computation is carried out by creating pairs of
anyons from the ground state, separating them far apart,
transporting individual anyons adiabatically around each other,
and finally fusing pairs of anyons together.
A list of particle types produced in the fusion is the classical
outcome of the computation.
An error rate of individual gates in TQC is expected to be
much smaller than $\delta$.

A physical system that may serve as a platform for TQC
is a two-dimensional electron gas in the Fractional Quantum Hall
Effect (FQHE) regime.
The FQHE plateau at the filling fraction
$\nu=5/2$ was observed by Willett et
al.~\cite{Wil87a} in late eighties.
Shortly after that Moore and Read~\cite{MR91a} developed a theory
predicting that elementary excitations of
the  $\nu=5/2$ state
are non-Abelian anyons.
The corresponding braid group representation
was found by Nayak and Wilczek~\cite{NW96a}.
For the sake of brevity we shall refer to the anyons
existing in the $\nu=5/2$ state as {\it Ising anyons}
(their exchange statistics can be described by
monodromy of holomorphic correlation functions
of the 2D Ising model~\cite{MR91a}).

From the experimental point of view, Ising anyons
have many favorable properties.
A large quasiparticles gap
(estimated as $\Delta \ge 100~\mbox{mK}$ in~\cite{Pan01a})
suppresses thermal creation of `stray' particles, while
non-zero electric charge  permits control of anyons using
electrostatic gates. Besides, one can take advantage
of the well-developed FQHE experimental technology.
An experimental setup for controlling Ising anyons and
testing their statistics has been recently
proposed by several authors~\cite{FNTW98a,Sar05a,Bon05a,Hal05a}.
An error rate for the one-qubit $\sigma^x$ operation has
been estimated as $10^{-30}$ in~\cite{Sar05a}.

The only fact that prevents one from using Ising anyons for TQC is that
the braid group
representation describing their statistics is  not computationally
universal.
We shall see that one can easily compute an amplitude of any braiding
process, see Section~\ref{sec:TQC}.
Loosely speaking, TQC with Ising anyons is an intersection of
two computational models known to be classically
simulatable: quantum circuits with
Clifford gates~\cite{Got98s,CN00a,AG04s},
and Fermionic Linear Optics~\cite{TD02a,Kni01f,Bra05a}.
Therefore, Ising anyons offer only
reliable storage of quantum information  and reliable implementation
of a certain non-universal gate set, see Section~\ref{sec:TQC} for details.

The goal of the present paper is to argue that this drawback is not
as serious as it might seem. We show that a universal gate set can be
simulated by standard TQC operations, i.e.,  adiabatic
transport and fusion of anyons, and very noisy non-topological
operations, such as direct short-range interaction of anyons.
The latter can be thought of as a tunneling process in which two anyons
exchange a virtual quasiparticle. It can be implemented by
transporting two anyons sufficiently close to each other,
waiting for an appropriate period of time, and then returning the
anyons to the original positions. Another example of a
non-topological computational element is a two-point contact interferometer
proposed in~\cite{Bon05a,Hal05a,FNTW98a}. It has the geometry of a Hall bar
with two constrictions, such that
quasiparticle tunneling occurs between two edge currents
on the opposite edges of the bar.
The tunneling current is sensitive to
the total topological charge of anyons trapped inside the
interferometer loop.
We will show that the short-range interaction and the two-point
contact interferometer together with TQC operations provide
a universal gate set.

Our main result concerns the threshold error rate of non-topological
operations. To avoid propagation of errors we apply all non-topological
operations before the computation itself to prepare a supply
of ``computationally universal'' ancillary states from the vacuum.
In our scheme there will be two types of ancillary states:
a four-particle state $|a_4\ra$
and an eight-particle state $|a_8\ra$.
From the computational perspective, $|a_4\ra$ can be identified
with a one-qubit state
$2^{-1/2}(|0\ra + e^{i\pi/4}|1\ra)$
(we represent a qubit by four quasiparticles).
Analogously, $|a_8\ra$ can be identified with a two-qubit state
$2^{-1/2}(|0,0\ra + |1,1\ra)$.
One copy of $|a_4\ra$ together with TQC
operations allows the implementation of the one-qubit $\pi/8$-rotation.
Analogously, one copy of $|a_8\ra$ allows the implementation of the CNOT gate.
Summarizing, universal computation can be carried out by TQC
operations if a supply of states $|a_4\ra$ and $|a_8\ra$ is available.

Since non-topological operations are not perfect,
in practice one can prepare only some very noisy
ancillary states $\rho_4$ and $\rho_8$ approximating $|a_4\ra$ and
$|a_8\ra$ up to some precision. We characterize this
precision by two parameters
\[
\ep_4=1-\la a_4|\rho_4|a_4\ra \quad \mbox{and}
\quad \ep_8 = 1-\la a_8|\rho_8|a_8\ra.
\]
We prove that the ideal states $|a_4\ra$ and $|a_8\ra$ can be distilled from
many copies of $\rho_4$ and $\rho_8$ by TQC operations
provided that
\begin{enumerate}
\item All TQC operations are perfect,
\item $\ep_4<0.14$,
\item $\ep_8<0.38$.
\end{enumerate}
A distillation method that we use is a combination of
``magic states distillation'' proposed in~\cite{BK04a}
and a slightly modified version of
the entanglement purification protocol of Bennett et
al.~\cite{Ben95a,BDSW96a}.

Summarizing, if
one can prepare the states $|a_4\ra$, $|a_8\ra$ accurately enough,
such that the conditions above are satisfied,
then any quantum computation can be efficiently
simulated by Ising anyons.
The overall simulation requires only poly-logarithmic overhead.
Specifically, the number of noisy ancillas $\rho_4$, $\rho_8$,
and the number of TQC operations that one needs to simulate
a quantum circuit with $L$ gates scales as
$L\, (\log L)^3$.

In the case when one can meet only the condition $\ep_8<0.38$, TQC operations
allow one to implement any Clifford gates
(i.e., the CNOT gate, the Hadamard gate,
and the one-qubit $\pi/4$-rotation). Though these gates do not constitute a
universal set, they are sufficient to implement any error correction
scheme based on stabilizer codes~\cite{KSV02a}. Error correction
might be needed if one takes into account finite error rate of TQC
operations (which is neglected throughout this paper).

Our derivation of the threshold conditions~2,3 is based on a single
assumption regarding the error model
characterizing non-topological operations --- they must
obey the superselection rules of Ising anyons.
Accordingly, we assume that matrix elements of $\rho_4$ and $\rho_8$
are non-zero only for the vacuum sector (recall that each ancilla
is prepared from the vacuum).

The rest of the paper is organized as follows.
Section~\ref{sec:Ising} provides the necessary background on Ising
anyons. In Section~\ref{sec:TQC} a TQC with Ising anyons is
discussed and its classical simulatibility is proved.
Section~\ref{sec:a8} describes a distillation method for the state
$|a_8\ra$. We show how to use ancillas $|a_8\ra$ to implement
Clifford group gates in Section~\ref{sec:Clifford}.
Finally, in Section~\ref{sec:UQC} we make use of  the magic states
distillation protocol to simulate universal computation.
Also the efficiency of the simulation is analyzed.
Some particular non-topological
ancilla preparation methods are discussed in Section~\ref{sec:nto}.

\section{Ising anyons}\label{sec:Ising}

A complete specification of any class of anyons is rather
complicated and involves a lot of data
including a list of particle types, their fusion and braiding rules,
$S$-matrices, e.t.c., see~\cite{Kit05l} for a comprehensive
review, and~\cite{Bon05a} for a detailed discussion
of Ising anyons in the context of FQHE.
In this section we briefly outline the properties of Ising anyons
focusing on those relevant for quantum computation.

\subsection{Particle types and fusion rules}

There are two non-trivial particle types in the class of
Ising anyons. We shall label them by letters $\ssigma$ and
$\ppsi$. Particles of different type
cannot be converted to one another (or to
the vacuum) by a local operator, thus describing superselection
sectors of the model.
However, if one brings two particles close to each other, they
can fuse into a single one, or annihilate each other
forming a topologically trivial particle (belonging to the
vacuum sector). Admissible
interconversions of particles are formally described by
fusion rules:
\be\label{fusion}
\ppsi \times \ppsi = \one,
\quad
\ppsi \times \ssigma = \ssigma,
\quad
\ssigma\times \ssigma = \one + \ppsi.
\ee
Here $\one$ stands for the vacuum sector.
The most important for us is the last  rule.
It implies that a pair of $\ssigma$-particles
can be prepared in two orthogonal states
that differ by the total topological charge.
Computing a product $\sigma\times
\cdots\times \sigma$ for
$2n$ $\ssigma$-particles one can easily get
\be\label{2n_fusion_rule}
\ssigma^{\times\, 2n} = 2^{n-1} \, \one + 2^{n-1}\, \ppsi.
\ee
Thus if one creates $2n$ $\ssigma$-particles from the vacuum,
there is a $2^{n-1}$-dimensional subspace of states
that can be distinguished by fusing some pairs of particles
together and observing a type of the resulting particles.
It is used as a computational space in the TQC scheme.

\subsection{The braid group representation}

\begin{center}
\begin{figure}
\includegraphics[scale=0.3]{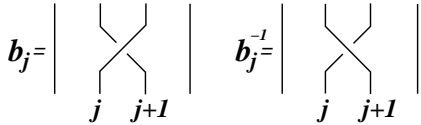}
\caption{Braid group generators.}\label{fig:braids}
\end{figure}
\end{center}

Recall that exchange statistics of particles living in the
$2+1$ dimensional space-time is described by unitary representations
of the braid group rather than the symmetric group
(because the clockwise and the counterclockwise exchanges
are not equivalent).
The braid group $\calB_n$ with $n$ strings can be formally
described by generators $b_j$, $b_j^{-1}$,
$j=1,\ldots,n-1$ (see Fig.~\ref{fig:braids})
that obey the Yang-Baxter relations
\begin{eqnarray}
b_j\, b_k = b_k\, b_j &\mbox{for}&
|j-k|>1, \nn \\
b_j \, b_{j+1}\, b_j = b_{j+1}\, b_j\, b_{j+1},
&\mbox{for} &
j=1,\ldots,n-2.
\end{eqnarray}
The strings can be thought of as world lines of particles,
whose initial and final positions are chosen on the $x$-axis.

Exchange statistics of $\sigma$-particles is
described by the {\it spinor representation} of the braid
group~\cite{NW96a}
\[
\varphi\, : \, \calB_{2n} \to U(2^n).
\]
It is constructed using the spinor representation
of the orthogonal group $SO(2n)$. Let us introduce
the Pauli operators $\sigma^x_j$,
$\sigma^y_j$, $\sigma^z_j$ on $n$ qubits, and auxiliary {\it Majorana operators}
$\cop_1,\cop_2,\ldots,\cop_{2n}$ defined as
\begin{eqnarray}\label{Jordan-Wigner}
\cop_{2j-1}&=&\sigma^z_1\otimes \cdots \otimes \sigma^z_{j-1}\otimes
\sigma^x_j\otimes I_{j+1} \otimes \cdots \otimes I_n,  \nn \\
\cop_{2j}&=&\sigma^z_1\otimes \cdots \otimes \sigma^z_{j-1}\otimes
\sigma^y_j\otimes I_{j+1} \otimes \cdots \otimes I_n,  \nn \\
\end{eqnarray}
where $I$ stands for the one-qubit identity operator
and $j$ runs from $1$ to $n$.
The Majorana operators obey commutation rules
\[
\cop_p\, \cop_q + \cop_q\, \cop_p=2\delta_{pq}\, I, \quad
\cop_p^\dag=\cop_p,
\quad \mbox{for any} \quad p,q.
\]
Then the spinor representation of the braid group generators
$b_1,\ldots,b_{2n-1}$ is defined as
\[
\varphi(b_p) = \exp{\left(-\frac{\pi}4\,  \cop_p\, \cop_{p+1} \right)}=
\frac1{\sqrt{2}} \left( I -\cop_p\, \cop_{p+1} \right).
\]
(We have omitted the overall phase of
$\varphi(b_p)$, since it is irrelevant for quantum computation purposes.)

The Yang-Baxter relations can be easily verified using the following identity:
\[
\varphi(b_p) \, \cop_q \, \varphi(b_p)^\dag =
\left\{ \ba{rcl}
\cop_q, &\mbox{if} & q\notin \{p,p+1\}, \\
\cop_{p+1}, &\mbox{if} & q=p, \\
-\cop_p, &\mbox{if} & q=p+1.\\
\ea
\right.
\]
It says that an exchange of adjacent $\sigma$-particles is equivalent
to an exchange of the corresponding Majorana operators (up to a sign).

\subsection{Topological charge measurements}

The multi-dimensionality of the braid group representation accounts
for the fact that there is more than one way to fuse $2n$
$\sigma$-particles into the vacuum (or into $\ppsi$-particle).
A process in which two adjacent $\sigma$-particles $p$ and $p+1$
are fused together and then a type of the resulting particle
($\one$ or $\ppsi$) is observed can be described as a
projective measurement of an observable
\[
F_p=-i\, \cop_p\, \cop_{p+1}.
\]
The eigenvalues $+1$ and $-1$ correspond to the
resulting particle's type $\one$ and $\ppsi$
respectively.

A type of a particle that one would obtain by fusing together
all $2n$ $\sigma$-particles is measured by a {\it parity operator}
\be\label{parity}
Q=\sigma^1_1\otimes \cdots \otimes \sigma^z_n = (-i)^n\,
\cop_1\cop_2\cdots \cop_{2n}.
\ee
Note that $Q$ commute with the action of any braid group
element, as well as with observables $F_p$.
This is a manifestation of the superselection rules ---
any local operator preserves the total topological charge.
Any state $|\Psi\ra$ of $2n$ $\sigma$-particles that can be
created from the vacuum obeys $Q\, |\Psi\ra = + |\Psi\ra$.
Analogously,  $|\Psi\ra$ can be prepared starting from
a single $\ppsi$-particle iff $Q\, |\Psi\ra = - |\Psi\ra$.

{\it Remark:} Strictly speaking,
fusion is a process reducing the Hilbert space of states,
since it replaces two particles by one.
To simplify the notations we describe fusion as a projective
measurement. This is justified, since a fusion can always
be followed by an auxiliary fission process in which the resulting
$\one$ or $\ppsi$ particle is split into a pair of
$\sigma$-particles.

\section{Topological Quantum Computation with Ising anyons}\label{sec:TQC}

The goal of this section is to introduce a computational
model that captures all features of TQC with Ising anyons.
We will show that any computation within this model can be
efficiently simulated classically. Finally, we describe
a natural encoding of a qubit by $\sigma$-particles.

\subsection{Formal computational model}
To define a formal model we just need to extract
its constituents
from Section~\ref{sec:Ising} ---
the computational Hilbert space with a fiducial
initial state, a set of unitary gates, and a set
of admissible measurements.

The computational Hilbert space of $n$ qubits
\[
\calF_n = (\CC^2)^{\otimes n}
\]
will be represented by $2n$ $\sigma$-particles.
The initial state $|\zero\ra=|0\ra\otimes \cdots \otimes |0\ra$
is prepared by preparing
pairs of $\sigma$-particles $(1,2),\ldots, (2n-1,2n)$ from the vacuum.

A set of elementary unitary gates includes nearest-neighbors
exchange operations
\[
B_p\equiv \varphi(b_p) =
\exp{\left(-\frac{\pi}4\,  \cop_p\, \cop_{p+1} \right)}.
\]
For any $p<q$ define a non-local exchange operation
\be\label{braid_gate}
B_{p,q}=\exp{\left(-\frac{\pi}4\,  \cop_p\, \cop_{q} \right)}.
\ee
Its conjugated action on the Majorana operators is
\be\label{braid_gates_conjugated}
B_{p,q} \, \cop_r \, B_{p,q}^\dag =
\left\{ \ba{rcl}
\cop_r, &\mbox{if} & r\notin \{p,q\}, \\
\cop_{q}, &\mbox{if} & r=p, \\
-\cop_p, &\mbox{if} & r=q.\\
\ea
\right.
\ee
One can easily verify that a non-local exchange is
a composition of $O(n)$ nearest-neighbor exchanges,
namely for any $p\le q-2$ one has
\[
B_{p,q}= B_{q-1} \cdots B_{p+1} \, B_p \, B_{p+1}^\dag \cdots
B_{q-1}^\dag.
\]
The operations $B_{p,q}$ constitute a set of elementary
unitary gates in our model. We shall refer to them as
{\it braid gates}.

Finally, a set of measurements includes
nearest-neighbors two-particle fusion processes, i.e.,
non-destructive
projective measurements of observables $F_p=-i\, \cop_p\, \cop_{p+1}$.
For any $p<q$ define an observable
\[
F_{p,q}=-i\, \cop_p\, \cop_q.
\]
Taking into account that $F_{p,q}=B_{p+1,q}\, F_p\, B_{p+1,q}^\dag$,
we can also measure eigenvalue of any observable $F_{p,q}$.
Summarizing, the formal computational model is as follows:
\begin{itemize}
\item The Hilbert space: $\calF_n=(\CC^2)^{\otimes n}$,
\item The initial state: $|\zero\ra=|0\ra\otimes \cdots \otimes |0\ra$,
\item Braid gates: $B_{p,q}=\exp{\left(-\frac{\pi}4\,  \cop_p\,
\cop_{q} \right)}$,
\item Measurable observables: $F_{p,q}=-i\, \cop_p\, \cop_q$.
\end{itemize}
We shall refer to this list as a TQC model.
It will be assumed throughout this paper that TQC operations
are implemented perfectly (a storage of quantum states
in $\calF_n$ is also assumed to be perfect).

\subsection{Classical simulation of TQC with Ising anyons}

The fact that any computation in the TQC model can be simulated
classically follows easily from the Gottesman-Knill
theorem, see~\cite{CN00a}.
Indeed, taking into account the relation  Eq.~(\ref{Jordan-Wigner})
between  the Pauli matrices and the Majorana operators,
and the conjugated action of the braid gates
Eq.~(\ref{braid_gates_conjugated}), one can easily prove
that any braid gate maps Pauli operators to Pauli operators
under a conjugation. Thus all braid gates belong to the Clifford group.
Since the set of measurable observables includes only
Pauli operators, we can directly apply the
stabilizer formalism~\cite{Got98s,AG04s} to simulate the TQC.

Another way to deduce the same result is
to relate the TQC model and the Fermionic Linear Optics (FLO),
see~\cite{Kni01f,TD02a,Bra05a}.
A theorem proved in these papers asserts that any computation
within the FLO model can be efficiently simulated classically.
In terms of FLO operations,
the initial state $|\zero\ra$ is the Fock vacuum,
the braid gates Eq.~(\ref{braid_gate})
are just special case of Bogolyubov canonical transformations,
while the observables $F_{p,q}$ measure
single-mode occupation numbers. Then the classical simulatibility
of the TQC model follows directly from~\cite{Bra05a}.

Loosely speaking, TQC with Ising anyons is an intersection
of two computational models known to be
classically simulatable:  the Clifford group/stabilizer
formalism model, and the FLO.
This is the reason why we need two types of ``computationally
universal'' ancillary states. The ancilla $|a_4\ra$
takes us beyond the Clifford group model, while
the ancilla $|a_8\ra$ introduces a non-linearity
necessary to go beyond the FLO model.

In the remainder of this subsection we explicitly describe
a set of unitary operators and a set of quantum states
that can be achieved by TQC operations.

Let $G \subset U(2^n)$ be a group generated by braid gates $B_{p,q}$
for $2n$ $\sigma$-particles.
To describe $G$ note that a subgroup $H\subset G$ generated by
double exchanges
$B_p^2=-\cop_p\, \cop_{p+1}$, $p=1,\ldots,2n-1$,
coincides with the set of all {\it even} products of Majorana operators
(we do not care about the overall phase of
operators). Thus if one parameterizes a product of Majorana operators
$\cop_1^{x_1}\cdots \cop_{2n}^{x_{2n}}$ by a binary $2n$-bit string
$(x_1,\ldots,x_{2n})$, we get
$H\cong (\ZZ_2)^{2n-1}$. Moreover, the subgroup $H$ is normal:
$B_p H B_p^\dag =H$ for any $p$.
One can easily check that the factor group $G/H$ coincides with the permutation group $S_{2n}$ of $2n$
objects. Thus $G$ can be represented as a semidirect product:
\[
G = (\ZZ_2)^{2n-1} \ltimes S_{2n}.
\]

To characterize the set of quantum states that can be prepared by
TQC operations, note that the initial state $|\zero\ra$ is a
stabilizer state with a stabilizer group
\[
S=(\cop_1\cop_2, \cop_3\cop_4,\ldots,\cop_{2n-1}\cop_{2n}).
\]
Applying any sequence of braid gates $B_{p,q}$ to this state
is equivalent to updating the stabilizer group
according to Eq.~(\ref{braid_gates_conjugated}).
A new stabilizer group is
\be\label{paired_stabilizer_group}
S'=(\cop_{p(1)}\cop_{p(2)}, \cop_{p(3)}\cop_{p(4)}, \ldots,
\cop_{p(2n-1)}\cop_{p(2n)}),
\ee
where $p$ is a permutation of the numbers $\{1,2,\ldots,2n\}$.

Let $|\psi\ra$ be any state with a stabilizer group $S'$ as above.
A measurement of an observable $F_{p,q}$
has non-trivial effect on $|\psi\ra$ only if
$\cop_p\, \cop_q$ is not a stabilizer of $|\psi\ra$.
In this case $p$ and $q$ must belong to different pairs, i.e.,
$\cop_r\cop_p$ and $\cop_q\cop_s$ are stabilizers of $|\psi\ra$
for some integers $r\ne s$. Moreover, these are the only
generators of $S'$ that anticommute with $F_{p,q}$.
Therefore, measuring eigenvalue of $F_{p,q}$ is equivalent
to updating the stabilizer group according to
\[
(\ldots,\cop_r\cop_p,\cop_q\cop_s,\ldots)\to
 (\ldots,\cop_r\cop_s,\cop_p\cop_q,\ldots).
\]
We conclude that any state one can get from
the initial state $|\zero\ra$
by TQC operations can be described by
a stabilizer group Eq.~(\ref{paired_stabilizer_group}) for some
permutation $p\in S_{2n}$.

{\it Remark:} In the arguments above we have ignored eigenvalues
associated with stabilizer operators which may be either
$+i$ or $-i$.
 Naturally, after each
transformation one has to update the eigenvalues as well.
For simplicity we skip these details.

\subsection{Representation of a qubit}

So far we represented a single qubit
by a pair of $\sigma$-particles. Although this is the most
efficient representation in terms of resources,
it has some serious drawbacks.
Since a pair of $\sigma$-particles
prepared in the basis states $|0\ra$ and $|1\ra$ has
the total topological charge $\one$ and $\ppsi$ respectively,
a qubit cannot be prepared in a superposition of the basis states,
e.g., $|0\ra \pm |1\ra$, because
they violate the superselection rules.

For this reason we shall represent a logical qubit by
a group of four $\sigma$-particles. The basis states $|\bar{0}\ra$
and $|\bar{1}\ra$ of a logical qubit will be identified with
physical states
$|0,0\ra\in \calF_2$ and $|1,1\ra\in\calF_2$. Both these states have the trivial total charge.
A computational subspace spanned by $|0,0\ra$ and $|1,1\ra$ can be
specified by an eigenvalue equation
\be\label{1234}
-\cop_1\cop_2\cop_3\cop_4 \, |\psi\ra=|\psi\ra.
\ee
Logical Pauli operators
$\bar{\sigma}^x$, $\bar{\sigma}^y$, and
$\bar{\sigma}^z$ acting on the computational subspace can be chosen as
\begin{eqnarray}
\bar{\sigma}^z &=& -i\cop_1\cop_2,\nn \\
\bar{\sigma}^x &=& -i\cop_2\cop_3, \nn \\
\bar{\sigma}^y &=&  -i \cop_1\cop_3.
\end{eqnarray}
Clearly, logical Pauli operators can be implemented by braid gates, for
example, $\bar{\sigma}^z$ corresponds to  winding the particle $1$
around the particle $2$. Besides, TQC operations allow one
to measure an eigenvalue of the
logical one-qubit Pauli operators.

Note that any four-particle braid gate commutes with the parity
operator $\cop_1\cop_2\cop_3\cop_4$, i.e., it implements some logical one-qubit gate.
To find a subgroup of $U(2)$ generated by these gates, it suffices to
consider the braid gates $B_{1,2}$, $B_{2,3}$, $B_{3,4}$. In terms of
logical Pauli operators one has
\[
B_{1,2}=B_{3,4}=\exp{ (-i\frac{\pi}4\, \bar{\sigma}^z )}=
e^{-i\frac{\pi}4} \left( \ba{cc} 1 & 0 \\ 0 & i \\ \ea \right),
\]
\[
B_{2,3}=\exp{ (-i\frac{\pi}4\, \bar{\sigma}^x )}=
\frac1{\sqrt{2}} \left( \ba{cc} 1 & -i \\ -i & 1 \\ \ea \right).
\]
(By abuse of notations, we identify a braid gate and the
corresponding logical operator.)
These gates generate the one-qubit Clifford group $Cl(1)\subset U(2)$.

The four-particle qubit representation also has some drawbacks which
come out if one considers two logical qubits. Let us show that
any two-qubit logical state
\[
|\psi\ra=a\, |\bar{0},\bar{0}\ra + b\, |\bar{0},\bar{1}\ra + c\,
 |\bar{1},\bar{0}\ra
+  d\, |\bar{1},\bar{1}\ra \in \calF_4
\]
that can be prepared by TQC operations
has a product form:
\[
|\psi\ra=|\psi_1\ra\otimes |\psi_2\ra.
\]
Here $|\psi_1\ra$ and $|\psi_2\ra$ are some logical one-qubit states.
Indeed, we already know that $|\psi\ra$ obeys stabilizer equations
\be\label{12_34_56_78}
\cop_{p(1)}\cop_{p(2)}\, |\psi\ra = \pm i\, |\psi\ra,
\ldots,
\cop_{p(7)}\cop_{p(8)}\, |\psi\ra = \pm i\, |\psi\ra,
\ee
for some permutation $p\in S_8$, see
Eq.~(\ref{paired_stabilizer_group}).
On the other hand, the assumption that $|\psi\ra$ is a logical
two-qubit state implies that
\be\label{1234_5678}
-\cop_1\cop_2\cop_3\cop_4\, |\psi\ra=-\cop_5\cop_6\cop_7\cop_8\,
 |\psi\ra= |\psi\ra.
\ee
Obviously, Eq.~(\ref{12_34_56_78}) and
Eq.~(\ref{1234_5678}) are consistent with each other iff
for any $1\le j\le 4$ one has
\[
p(2j-1),p(2j) \in \{1,2,3,4\},
\]
or
\[
p(2j-1),p(2j)\in \{5,6,7,8\}.
\]
In other words, each stabilizer $\cop_{p(2j-1)}\cop_{p(2j)}$ of the
state $|\psi\ra$ is composed either from generators
$\cop_1,\cop_2,\cop_3,\cop_4$,
or from the generators $\cop_5,\cop_6,\cop_7,\cop_8$. It means that
$|\psi\ra$ has a product structure $|\psi\ra=|\psi_1\ra\otimes |\psi_2\ra$.

Summarizing, the four-particle qubit representation
allows one to prepare qubits in a superposition, but
no entangled states can be prepared topologically.

\noindent
{\bf No-entanglement rule:} The only logical states that can be
prepared by TQC operations from the initial state $|\zero\ra$
are products of one-qubit states.

\section{Purification of the eight-particle ancillas}
\label{sec:a8}

One way to get around the no-entanglement rule
is to use some very noisy non-topological operations
to prepare a state $\rho$ that
approximates some logical entangled `target' state.
Then one can try to improve accuracy of the approximation
by running a purification protocol
involving only TQC operations.
This is the strategy that we shall follow in this section.

\subsection{Outline}
A target state which we would like to purify is
the maximally entangled two-qubit logical state
\be\label{psi8}
|a_8\ra=\frac1{\sqrt{2}}(|\bar{0},\bar{0}\ra
+ |\bar{1},\bar{1}\ra)=\frac1{\sqrt{2}}(|0,0,0,0\ra + |1,1,1,1\ra).
\ee
It consists of eight $\sigma$-particles~\footnote{Under certain
natural assumptions, eight is the minimal number
of $\sigma$-particles
one has to start with to prepare
an entangled logical state.
The reason is that any state $|\psi\ra\in \calF_k$, $k\le 3$,
with a trivial total charge
is a Gaussian fermionic state, see~\cite{Bra05b} for a proof.
The arguments used to prove the no-entanglement rule can be easily
generalized to any Gaussian state since it also possesses a paired
structure. }
The quasiparticles $1,2,3,4$ and $5,6,7,8$ represent
the first and the second logical qubit respectively.

Let us denote $D(\calH)$ the set of all (mixed) quantum states on the
Hilbert space $\calH$.
Let $\rho\in D(\calF_4)$ be
eight-particle  mixed state that we can prepare by non-topological operations.
A precision up to which $\rho$ approximates $|a_8\ra$
can be characterized by a parameter
\[
\ep=1-\la a_8|\rho|a_8\ra.
\]
It will be referred to as an {\it error rate}.

The only assumption we made about $\rho$ is that
it has a support only on the even subspace
of $\calF_4$, i.e.,
\be\label{Q=1}
Q\, \rho = \rho\, Q =\rho,
\ee
where $Q$ is the total parity operator,
\[
Q=\cop_1\cop_2\cop_3\cop_4\cop_5\cop_6\cop_7\cop_8 = \sigma^z_1\otimes
\sigma^z_2\otimes \sigma^z_3\otimes \sigma^z_4.
\]
This assumption is justified if $\rho$ is prepared starting from the
vacuum by a local operator.
As was mentioned in Section~\ref{sec:Ising}, the operator $Q$
measures the total topological charge ($\one$ or $\ppsi$),
so Eq.~(\ref{Q=1}) is a consequence of the superselection rules.

An orthogonal projector onto $|a_8\ra$ looks as
\[
|a_8\ra\la a_8| = \frac1{16}
(I+S_1)(I+S_2)(I+S_3)(I+Q),
\]
where
\begin{eqnarray}\label{Ss}
S_1 &=& -\cop_1\cop_2\cop_5\cop_6, \nn \\
S_2 &=& -\cop_2\cop_3\cop_6\cop_7, \nn \\
S_3 &=& -\cop_1\cop_2\cop_3\cop_4.
\end{eqnarray}
Given a binary string $\ss=(s_1,s_2,s_3)$, $s_j\in\{0,1\}$,  consider
a normalized vector $|\Psi_{\ss}\ra\in \calF_4$ such that
\[
S_j \, |\Psi_\ss\ra = (-1)^{s_j}\, |\Psi_s\ra, \quad j=1,2,3,
\quad Q\, |\Psi_\ss\ra = |\Psi_\ss\ra.
\]
Notice that $|a_8\ra=|S_{000}\ra\equiv |S_{\bf 0}\ra$.
Obviously, $\{|\Psi_\ss\ra\}$ constitute an orthonormal basis
of the even subspace of $\calF_4$.
Therefore, $\rho$ can be written as
\be\label{original_ancilla}
\rho = \sum_{\ss,\tt} \rho_{\ss\tt} |\Psi_\ss\ra\la \Psi_\tt|,
\quad \rho_{{\bf 00}}=1-\ep.
\ee
By analogy with quantum error correcting codes,
the operators $S_j$ and the string $\ss$
will be referred to as {\it stabilizers}
and a {\it syndrome} respectively.

The goal of a purification is  to prepare one copy
of $|a_8\ra$ with an arbitrarily
small error rate $\ep'$ starting from
$n$ noisy copies of $|a_8\ra$ with an error rate
$\ep$. Performance of a purification protocol
can be characterized by a threshold value of $\ep$
below which the purification is possible, and
efficiency, i.e.,
an asymptotic behavior of $n=n(\ep,\ep')$ for $\ep'\to 0$.
We shall describe a protocol for which
the threshold error rate is
\be\label{a8:threshold}
\delta_8\approx 0.384.
\ee
and
\be\label{a8:efficiency}
n(\ep,\ep')\approx C\, ( -\log \ep'  )^3
\ee
for any fixed $\ep<\delta_8$ and $\ep'\to 0$.
Here $C$ is a function of $\ep$ only.
The protocol succeeds with a probability
at least $1/2$ and there is a flag that tells us
when it fails.

The protocol involves the following steps:
\begin{itemize}
\item Dephasing: make $\rho$ diagonal in the basis
$\{|\Psi_\ss\ra\}$;
\item Syndrome whirling: make the probability distribution of
the non-zero syndromes $\ss\ne 0$ uniform;
\item Purification: convert two noisy copies of $|a_8\ra$
into one clean copy by postselective measurements on
four pairs of $\sigma$-particles.
\end{itemize}
In order to achieve an arbitrarily small error rate,
these steps have to be repeated sufficiently many times in a recursive fashion.
Below we describe the protocol on a more technical level.

\subsection{Dephasing}
Let $S$ be a group generated by $S_1,S_2,S_3$.
It consists of eight elements.
Consider a quantum operation
\[
\Phi_S(\rho)=\frac18 \sum_{U\in S} U\rho\, U^\dag.
\]
It symmetrizes a state over the group $S$, thus
implementing a dephasing in the basis $\{|\Psi_\ss\ra\}$.
The stabilizer operators $S_j$ themselves
can be implemented using  braid gates,
for instance, $S_1=-B_{1,2}^2 B_{5,6}^2$.
Accordingly, $\Phi_S$ can be implemented using
braid gates, if $U\in S$ is drawn randomly according to
the uniform distribution.
Obviously, for any state $\rho$ one has
\be\label{sym_ancilla}
\Phi_S(\rho)=\sum_\ss p(\ss) \, |\Psi_\ss\ra\la \Psi_\ss|,
\quad p(\ss)\equiv \la \Psi_\ss|\rho|\Psi_\ss\ra.
\ee
We shall assume that each ancilla is
acted on by $\Phi_S$ before it is fed into the purification
protocol.
It allows one to identify quantum states with
probability distributions of syndromes.

\subsection{Syndrome whirling}
A probability distribution of syndromes $p(\ss)$ can be brought
by braid gates into
the standard bimodal form
\be\label{standard_p}
p(\ss)=\left\{ \ba{rcl} 1-\ep &\mbox{if} & \ss=(0,0,0),\\
                        \ep/7 &\mbox{if} & \ss\ne (0,0,0).\\
\ea\right.
\ee
To achieve this,
we will firstly show how to
implement a cyclic shift on the set of seven non-zero syndromes
$\ss\ne {\bf 0}$.
Then we shall implement a random cyclic shift.

Consider a braid-gate
\be
U_{12}=B_{2,3}\, B_{6,7}^\dag.
\ee
Its conjugated action is as follows
(only non-trivial part of the action is shown):
\[
U_{12}\cdot U_{12}^\dag = \left\{ \ba{rcl} \cop_2 &\to & \cop_3, \\
                          \cop_3 &\to & -\cop_2, \\
              \cop_6 &\to & -\cop_7, \\
              \cop_7 &\to & \cop_6. \\
\ea \right.
\]
Accordingly, a  conjugated action of $U_{12}$ on the stabilizers
$S_j$ is
\[
U_{12}\cdot U_{12}^\dag = \left\{ \ba{rcl}
 S_1 &\to & S_1\, S_2, \\
S_2 &\to & S_2, \\
S_3 & \to & S_3.\\
\ea \right.
\]
Therefore, $U_{12}$ implements
a XOR-like transformation
\[
U_{12}\, |\Psi_{s_1,s_2,s_3}\ra = |\Psi_{s_1\oplus s_2,s_2,s_3}\ra,
\]
where $\oplus$ stands for the addition by modulo $2$.
Analogously, one can check that braid gates
\[
U_{23}=B_{1,2}^\dag\, B_{3,4} \quad \mbox{and} \quad
U_{31}=B_{1,5}\, B_{2,6}
\]
implement XOR-like transformations
\begin{eqnarray}
U_{23}\, |\Psi_{s_1,s_2,s_3}\ra &=& |\Psi_{s_1,s_2\oplus s_3,s_3}\ra,\nn \\
U_{31}\, |\Psi_{s_1,s_2,s_3}\ra &=& |\Psi_{s_1,s_2,s_1\oplus s_3}\ra.
\end{eqnarray}
Consider now a braid gate
\[
U=U_{31} U_{23} U_{12}.
\]
One can easily check that $U$ implements a cyclic shift
of non-zero syndromes:
\[
U\, |\Psi_\ss\ra = |\Psi_{\eta(\ss)}\ra, \quad \eta=\left(
\ba{cccccccc} 0 & 1 & 2 & 3 & 4 & 5 & 6 & 7 \\
              0 & 3 & 7 & 4 & 5 & 6 & 2 & 1 \\
\ea \right).
\]
Here $\eta$ is a permutation of the numbers $\{0,1,\ldots,7\}$,
and the syndromes  are represented
by integers according to $\ss=4s_1 + 2s_2 + s_3$.
Consider a symmetrization $\Phi_U$ over the
cyclic group generated by $U$ (it is the cyclic group $\ZZ_7$,
since $U^7=I$), i.e.,
\[
\Phi_U(\eta)=\frac17 \sum_{p=0}^6 U^p\, \eta \, U^{-p}.
\]
An application of $\Phi_U$ to a state
$\rho=\sum_\ss p(\ss)\, |\Psi_\ss\ra\la\Psi_\ss|$
transforms the probability distribution $p(\ss)$ into
the standard form Eq.~(\ref{standard_p}), where
\[
\ep=1-p({\bf 0})=1-\la \Psi_{\bf 0}|\rho|\Psi_{\bf 0}\ra.
\]
By construction, $\Phi_U$ is a
probabilistic mixture of braid gates.

\subsection{Elementary purification round}

Suppose we are given a supply of states
\[
\rho=\sum_{\ss} p(\ss) \, |\Psi_\ss\ra\la \Psi_\ss|.
\]
Here $p(\ss)$ is some fixed  probability distribution of syndromes
(which may or may not have the standard bimodal form).

\begin{center}
\begin{figure}
\includegraphics[scale=0.4]{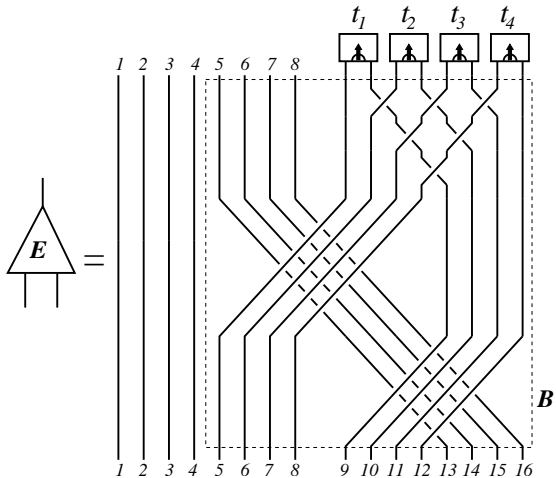}
\caption{The elementary purification round.}\label{fig:round}
\end{figure}
\end{center}

Consider a state $\rho\otimes \rho$ where
the first and the second copy
is composed from generators $\cop_1,\ldots,\cop_8$
and $\cop_9,\ldots,\cop_{16}$ respectively.
Let us reshuffle the generators by a braid gate $B$ shown on
Fig.~\ref{fig:round} (inside the dashed rectangle) and then
measure eigenvalues of four operators
\be\label{Ts}
\ba{rclrcl}
T_1 &=& -i\cop_{9}\cop_{10}, &  T_2 &=& -i\cop_{11}\cop_{12}, \\
T_3 &=& -i\cop_{13}\cop_{14}, & T_4 &=& -i\cop_{15}\cop_{16}.\\
\ea
\ee
Let $t_1,t_2,t_3,t_4\in \{0,1\}$ be the measurement outcomes, such that
$T_j$ has an eigenvalue $(-1)^{t_j}$.
Since the input state $\rho\otimes \rho$ is a probabilistic mixture of
pure states $|\Psi_\rr\ra\otimes |\Psi_\ss\ra$, it suffices to
analyze the effect of the braiding+measurement
operation on these input states.
For any string of outcomes $\tt=(t_1,t_2,t_3,t_4)$
consider the final (unnormalized) state
\[
|F_{\tt|\rr\ss}\ra=P_\tt\,B \, |\Psi_\rr\otimes \Psi_\ss\ra,
\]
where
\[
P_\tt=\frac1{16}\prod_{j=1}^4\left( I+(-1)^{t_j} \, T_j\right)
\]
is the projector corresponding to the outcomes $\tt$.
Taking into account an identity
\be\label{the_identity}
(\cop_5\cop_6\cop_7\cop_8)(\cop_9\cop_{10}\cop_{11}
\cop_{12})= B^\dag \, (T_1 T_2 T_3 T_4)\, B
\ee
and the fact that $|\Psi_\rr\ra$, $|\Psi_\ss\ra$ are even states,
we conclude that
\be\label{check_sum}
|F_{\tt|\rr\ss}\ra=0 \quad \mbox{unless} \quad
r_3\oplus s_3=t_1\oplus t_2 \oplus t_3 \oplus t_4.
\ee
Thus a bit
\[
t\equiv t_1\oplus t_2\oplus t_3\oplus t_4
\]
can be regarded
as a check
sum for the syndrome bits $r_3$ and $s_3$. If $t=0$
then either both syndrome bits are correct,
$r_3=s_3=0$, or both of them are wrong, $r_3=s_3=1$.
If the input state $\rho$ has a sufficiently small error
rate (the probability distribution $p(\ss)$ is
concentrated at $\ss={\bf 0}$),
the former possibility is more likely than
the latter one. As we shall see now,
one can enhance a probability of states with a correct
eigenvalue of $S_3$ by
discarding the final state whenever the outcome
$t=1$ is observed.

Indeed, suppose we have measured $t=0$ and
$|F_{\tt|\rr\ss}\ra\ne 0$.
After some algebra one gets
\be\label{F}
|F_{\tt|\rr\ss}\ra=|\Psi_\uu\ra\otimes |t_1,t_2,t_3,t_4\ra
\ee
(up to a normalization), where
\begin{eqnarray}\label{final_syndrom}
u_1 &=& r_1 \oplus s_1\oplus t_1\oplus t_2\oplus 1, \nn \\
u_2 &=& r_2 \oplus s_2\oplus t_2\oplus t_3\oplus 1, \nn \\
u_3 &=& r_3 =s_3.
\end{eqnarray}
Note that the syndrome bit $u_3$ depends only upon
$\rr$ and $\ss$, while $u_1$ and $u_2$ depend
also upon $\tt$. Let us apply
additional braid gates
\begin{eqnarray}
\cop_2\cop_3 &:&  |\Psi_{u_1,u_2,u_3}\ra \to |\Psi_{u_1\oplus 1,
u_2,u_3}\ra, \nn \\
\cop_1\cop_2 &:&  |\Psi_{u_1,u_2,u_3}\ra \to |\Psi_{u_1,
u_2\oplus 1,u_3}\ra. \nn
\end{eqnarray}
conditioned on bits $t_1\oplus t_2\oplus 1$ and
$t_2\oplus t_3\oplus 1$ respectively.
These braid-gates flip the bits $u_1$ and $u_2$,
so we get
\be\label{final_syndrom_1}
u_1 = r_1 \oplus s_1, \quad
u_2 = r_2 \oplus s_2, \quad
u_3 = r_3 =s_3.
\ee
(To avoid clutter, the additional braid gates are not shown
on Fig.~\ref{fig:round}.)
Summarizing, the output state of
the elementary purification round is $|\Psi_\uu\ra$ with
$\uu$ determined by Eq.~(\ref{final_syndrom_1}).

For the mixed input state  $\rho\otimes \rho$
the syndromes $\rr$ and $\ss$ are drawn from a product
distribution $p(\rr)p(\ss)$, so the output (normalized) state is
\[
\rho_{out}=\sum_{\uu} p_{out}(\uu)\, |\Psi_\uu\ra\la\Psi_\uu|,
\]
where
\be\label{flow}
p_{out}(\uu)=Z^{-1}\, \sum_{\rr,\ss} \Gamma_{\rr,\ss}^{\uu}\,
p(\rr)\, p(\ss),
\ee
and
\[
\Gamma_{\rr,\ss}^{\uu} = \delta_{r_3,u_3}\,  \delta_{s_3,u_3}\,
\delta_{r_1\oplus s_1,u_1}\, \delta_{r_2\oplus s_2,u_2}.
\]
Normalizing $\rho_{out}$ one gets
\[
Z=\sum_{\rr,\ss,\uu} \Gamma_{\rr,\ss}^{\uu}\,  p(\rr)\, p(\ss) =
\sum_{\rr,\ss} \delta_{r_3,s_3}\, p(\rr)\, p(\ss).
\]
Note that $Z$ is equal to the  probability to observe
$t=0$, i.e., a success probability of the elementary
purification round.

\subsection{The protocol}
\begin{center}
\begin{figure}
\includegraphics[scale=0.3]{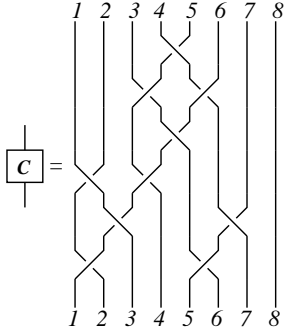}
\caption{A braid gate $C$ implementing a cyclic
shift of stabilizers $S_1\to S_2\to S_3\to S_1$.}\label{fig:cycle}
\end{figure}
\end{center}

Let $\ep^{(j)}$ and $\ep_{out}^{(j)}$ be a probability to
observe $s_j=1$ for the distribution $p(\ss)$ and
$p_{out}(\ss)$ respectively ($j=1,2,3$).
They can be regarded as error rates for the individual
syndrome bits.
If $p(\ss)$ has the standard bimodal form Eq.~(\ref{standard_p}), then
$\ep^{(j)}=4\ep/7$. On the other, for $\ep\ll 1$
one can easily find from Eq.~(\ref{flow}) that
\[
\ep_{out}^{(3)}\approx 16\ep^2/49,
\quad
\ep_{out}^{(1)}=\ep_{out}^{(2)}\approx 4\ep/7.
\]
It tells us that the
error rate $\ep^{(3)}$ is suppressed quadratically,
$\ep_{out}^{(3)}\approx (\ep^{(3)})^2$, while
the error rates $\ep^{(1)},\ep^{(2)}$ remain
practically unchanged, $\ep_{out}^{(1)}\approx \ep^{(1)}$,
$\ep_{out}^{(2)}\approx \ep^{(2)}$.
For this reason we shall
iterate the elementary purification
round shown on Fig.~\ref{fig:round} three times to
purify all three syndrome bits $s_1,s_2,s_3$.
The iterations are interlaced with
an additional braid gate $C$ which shifts the syndrome bits
cyclically, i.e.,
\[
C\, S_1 \, C^\dag = S_2,\quad
C\, S_2 \, C^\dag = S_3,\quad
C\, S_3 \, C^\dag = S_1.
\]
These equations can be satisfied if $C$
transforms the generators $\cop_1,\ldots,\cop_8$ according to
\[
C\, : \, \left\{
\ba{rclrcl}
\cop_1 &\to & \cop_6, & \cop_5 &\to & -\cop_7, \\
\cop_2 &\to & \cop_2, & \cop_6 &\to & \cop_3, \\
\cop_3 &\to & \cop_1, & \cop_7 &\to & -\cop_4, \\
\cop_4 &\to & \cop_5, & \cop_8 &\to & \cop_8. \\
\ea
\right.
\]
An explicit implementation of $C$ is shown
on Fig.~\ref{fig:cycle}.

\begin{center}
\begin{figure}
\includegraphics[scale=0.3]{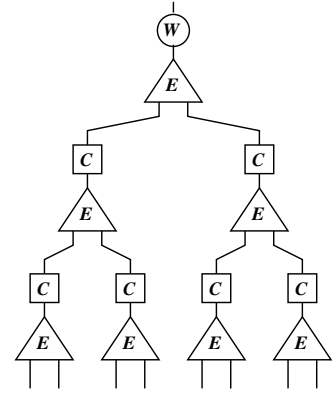}
\caption{A single round of
$a_8$-purification protocol. Time flows upwards.
Each line represents
one copy of the noisy $|a_8\ra$ state
(eight $\sigma$-particles).  Each triangle
$E$ corresponds to the elementary purification round
shown on Fig.~\ref{fig:round}.
Each rectangle $C$ represents a braid gate
that shifts the generators $S_1$, $S_2$, and $S_3$
cyclically, see Fig.~\ref{fig:cycle}.
Finally, a circle $W$ stands for the syndrome whirling
transformation.}\label{fig:a8}
\end{figure}
\end{center}

A single round of $a_8$-purification protocol is shown
on Fig.~\ref{fig:a8}.
Its input  consists of eight copies of a
noisy $|a_8\ra$ state in the
standard bimodal form with an error rate $\ep$:
\[
\rho=(1-\ep)\, |\Psi_{\bf 0}\ra\la \Psi_{\bf 0}|
+\frac{\ep}7 \, \sum_{{\bf s} \ne {\bf 0}}
|\Psi_\ss\ra\la \Psi_\ss|.
\]
The protocol outputs a single copy of
a noisy $|a_8\ra$ state
in the standard bimodal form with an error rate $\ep_{out}$.
The triangles labeled by `E' denote the elementary purification
rounds shown on Fig.~\ref{fig:round}.
The boxes labeled by `C' denote the braid gate
shown on Fig.~\ref{fig:cycle}. The circle labeled by `W'
stands for the syndrome whirling transformation.
Each line on the figure represents eight $\sigma$-particles.

\begin{center}
\begin{figure}
\includegraphics[scale=0.35,angle=-90]{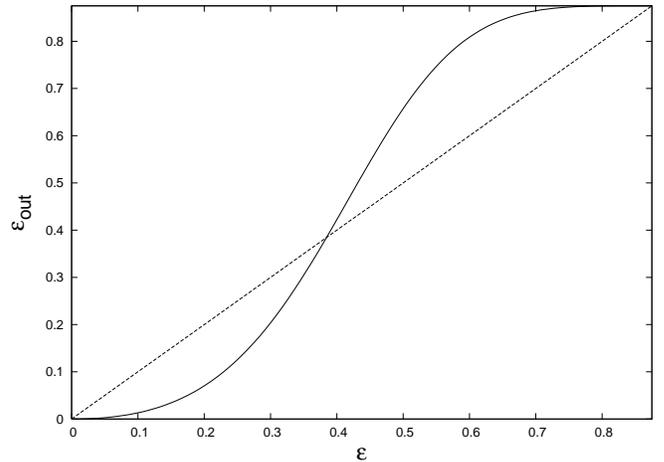}
\caption{Input/output error rates for
a single round of $a_8$-purification protocol.}
\label{fig:flow}
\end{figure}
\end{center}

The corresponding recursive flow equation $\ep_{out}(\ep)$
can be found by iterating Eq.~(\ref{flow}) three times
with an additional cyclic shifts inserted after each
iteration. Equivalently, $\ep_{out}(\ep)$ is
implicitly defined by equations
\begin{eqnarray}\label{flow_complete}
\ep_{out} &=& 1-Z^{-1}\, p_1({\bf 0}), \nn \\
p_1(\uu)&=& \sum_{\rr, \ss} \Theta_{\rr,\ss}^{\uu}
\,
p_2(\rr)\, p_2(\ss), \nn \\
p_2(\uu)&=&\sum_{\rr, \ss} \Delta_{\rr,\ss}^{\uu}
\,
p_3(\rr)\, p_3(\ss), \nn \\
p_3(\uu)&=&\sum_{\rr, \ss} \Gamma_{\rr,\ss}^{\uu}
\,
p(\rr)\, p(\ss), \nn \\
Z&=&\sum_{\uu} p_1(\uu).
\end{eqnarray}
The initial distribution $p(\ss)$ has the standard
bimodal form Eq.~(\ref{standard_p}) with the error rate
$\ep$. The coefficients $\Theta_{\rr,\ss}^{\uu}$
and $\Delta_{\rr,\ss}^{\uu}$ are obtained from
$\Gamma_{\rr,\ss}^{\uu}$ by a cyclic shift of indexes,
\[
\Theta_{\rr,\ss}^{\uu}=
\delta_{r_1,u_1}\,  \delta_{s_1,u_1}\,
\delta_{r_2\oplus s_2,u_2}\, \delta_{r_3\oplus s_3,u_3},
\]
\[
\Delta_{\rr,\ss}^{\uu}=
\delta_{r_2,u_2}\,  \delta_{s_2,u_2}\,
\delta_{r_1\oplus s_1,u_1}\, \delta_{r_3\oplus s_3,u_3}.
\]
The final cyclic shift can be discarded because it
is followed by the syndrome whirling.
We have found a solution of Eq.~(\ref{flow_complete}) using
MAPLE. A plot of a function $\ep_{out}(\ep)$
is shown on Fig.~\ref{fig:flow}.

The threshold error rate $\delta_8$ satisfying
$\ep_{out}(\delta_8)=\delta_8$ turns out to be
$\delta_8 \approx 0.384$.
If the initial error rate
is below the threshold, $\ep<\delta_8$, one can invoke the protocol
recursively to achieve arbitrarily small error rates.
For $\ep\ll 1$ one can easily get
\[
\ep_{out}(\ep)= \frac{48}{49}\ep^2 +O(\ep^3).
\]

A probability for all elementary purification rounds on
Fig.~\ref{fig:a8} to succeed is given by
normalizing coefficient $Z$ in Eq.~(\ref{flow_complete}).
For small $\ep$ one has
\be\label{Z}
Z=1-8\ep+O(\ep^2).
\ee
The function $Z(\ep)$ is monotone decreasing on the interval
$0\le \ep\le \delta_8$ and
$Z(\delta_8)\approx 0.04$.

\begin{center}
\begin{figure}
\includegraphics[scale=0.35,angle=-90]{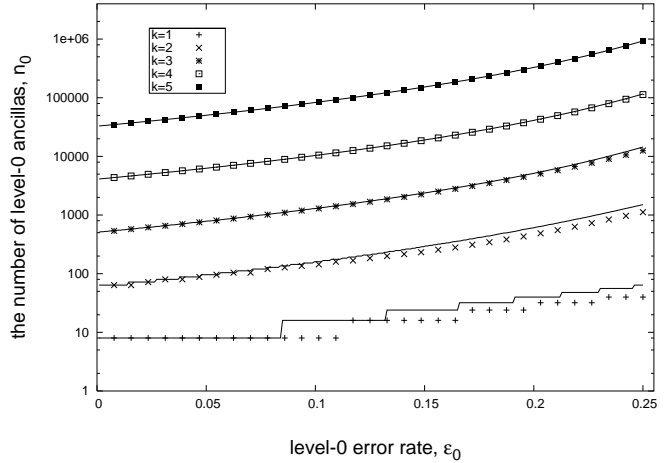}
\caption{The graph shows the number of
level-$0$ ancillas, $n_0$, that one needs to prepare one
level-$k$ ancilla  ($k=1,\ldots,5$)
with a success probability $1/2$.
The success probability has been evaluated using
Monte Carlo simulation of the protocol with $10^5$ trials.
Solid lines show the dependence $n_0(\ep_0)$
that one gets using a naive equation
$n_0=8^k \prod_{j=0}^{k-1} Z(\ep_j)^{-1}$
(it neglects fluctuations of $n_k$).}
\label{fig:MC}
\end{figure}
\end{center}

The initial supply of states $|a_8\ra$ with an error
rate $\ep_0\equiv \ep$ will be called level-$0$
ancillas.
Accordingly, level-$k$ ancillas are obtained from
the level-$0$ ancillas by iterating the protocol
shown on Fig.~\ref{fig:a8} $k$ times.
Let $\ep_k$ and
$n_k$ be an error rate and the total number of
level-$k$ ancillas. The numbers $\ep_{k+1}$, $n_{k+1}$
and $\ep_k$, $n_k$ are related by recursive flow equations
\be\label{naive_flow}
n_{k+1}\approx \frac{Z(\ep_k)}{8}\, n_k, \quad
\ep_{k+1}\approx \frac{48}{49}\, \ep_k^2.
\ee
Here fluctuations of the quantity $n_k$ are neglected.
Monte Carlo simulation of the $a_8$-purification protocol
shows that taking into account fluctuations does not
change the answer significantly, see Fig.~\ref{fig:MC}.

If one needs to prepare one copy of $|a_8\ra$ with
an error rate $\ep'$, the required number of levels $k$
can be found from an equation
\[
2^k\approx \frac{\log{(C\ep')}}{\log{(C\ep_0)}}, \quad C\equiv
\frac{48}{49}.
\]
The corresponding number of level-$0$ ancillas is
\be\label{naive}
n_0\approx 8^k \prod_{j=0}^{k-1} Z(\ep_j)^{-1}.
\ee
Assuming that $k\gg 1$ and denoting
$p(\ep_0)=\prod_{j=0}^{\infty} Z(\ep_j)$ (one can
easily check that this product is convergent), we get
\[
n_0\approx \frac{\log^3{(C\ep')}}{p(\ep_0)\log^3{(C\ep_0)}}.
\]
To find the probability for the protocol to convert
$n_0$ copies of the level-$0$ ancillas into one (or larger number)
level-$k$ ancilla, we used numerical simulations,
see Fig.~\ref{fig:MC}. The success probability
$P_s=\mbox{Prob}(n_k>0)$ was calculated as a function of $n_0$ using
Monte Carlo method.
For each particular $k=1,\ldots,5$  an equation $P_s(n_0)=1/2$
has been solved to find $n_0$ as a function of $\ep_0$.
As one can see from the figure,
the scaling of $n_0$ is pretty well described by Eq.~(\ref{naive}).

The operational cost of the purification,
i.e., the total number of braid gates and fusions
needed to achieve an error rate
$\ep'$ has the same scaling as $n_0$,
i.e., it is proportional to $(-\log{\ep'})^3$.

Suppose we have to prepare a
large number $L$ of ancillas $|a_8\ra$.
Let us first prepare $n=3L\cdot n_0(\ep_0,\ep')$ level-$0$ ancillas,
split them into $3L$ groups and then perform the $a_8$-purification
protocol independently in each group. If the purification succeeds in
each
group with a probability $1/2$,
the average number of
successful group is $3L/2$. Using the Chernoff bound one can
easily show that the probability for the number of successful group to
be smaller
than $L$ is at most $\exp{(-L/12)}$.
Thus one can say that a preparation of a single ancilla $|a_8\ra$
with an accuracy $\ep'$ costs about $(-\log \ep')^3$ elementary
operations (TQC operations and preparations of $\rho_8$).

This observation also shows that the purification can be described by
a trace preserving completely positive linear map
(in the exponentially rare events when the purification fails,
one can output an arbitrary state). Accordingly, we can generalize all
above results to the case when the preparation of the level-$0$
ancillas is a stochastic process that outputs a state $\rho_\alpha$
with a probability $p_\alpha$, such that $\sum_\alpha p_\alpha\,
\rho_\alpha=\rho$.

\section{Implementation of the Clifford group gates}\label{sec:Clifford}

Having prepared a supply of clean ancillas $|a_8\ra$ one
can proceed to the next goal ---
implementation of  entangling two-qubit gates.
We shall now explain how to implement a  two-qubit
controlled $\sm^z$ gate
\[
\Lambda(\sm^z)\, : |\bar{a},\bar{b}\ra\to
(-1)^{ab}\, |\bar{a},\bar{b}\ra
\]
acting on the logical qubits.
Together with logical one-qubit Clifford gates which can
be implemented by braid gates, see Section~\ref{sec:TQC}, it will
allow us to execute any Clifford group computation
(on the level of logical qubits).
We shall need the following technical result.

\begin{lemma}
The following operations
can simulate one another with assistance of TQC operations:
\begin{enumerate}\label{lemma:non-linear}
\item[O1.] A preparation of $|a_8\ra$,
\item[O2.]  A non-destructive measurement of an observable
$\cop_p\cop_q\cop_r\cop_s$ (all four labels are distinct).
\item[O3.] A unitary gate $\exp{\left(i\frac{\pi}4
  \cop_p\cop_q\cop_r\cop_s
\right)}$,
\end{enumerate}
\end{lemma}
{\it Remarks:}
(i) It is meant that one copy of any operation can be
exactly simulated
by one copy of any other operation.
(ii) The operator $\cop_p\cop_q\cop_r\cop_s$ has eigenvalues
$\pm 1$, so $O2$ can be described by orthogonal projectors
$(1/2)(I\pm \cop_p\cop_q\cop_r\cop_s)$.
(iii) If $O1$ is not ideal, so that $|a_8\ra$
has an error rate $\ep$, then $O2$ and $O3$ can
be executed with an error probability $O(\ep)$.
(iv) Explicit simulation protocols are given in the
proof of the lemma.

The controlled $\sm^z$ can be easily reduced to $O3$.
Indeed, suppose the first qubit is encoded by $\cop_1,\ldots,\cop_4$,
while the second qubit is encoded by $\cop_5,\ldots,\cop_8$.
Then
$\Lambda(\sigma^z)=\exp{(i\frac{\pi}4 (I-\bar{\sm}^z_1)(I-\bar{\sm}^z_2))}$,
where $\bar{\sm}^z_j$ are the logical Pauli operators defined as
$\sm^z_1=-i\cop_3\cop_4$ and $\sm^z_2=-i\cop_5\cop_6$, see Section~\ref{sec:TQC}.
Therefore
\begin{eqnarray}\label{cZ}
\Lambda(\sigma^z)&=&e^{i\frac{\pi}4}
\exp{\left(-i\frac{\pi}4 \cop_3\cop_4\cop_5\cop_6\right)}\nn \\
&& {}\cdot \exp{\left(-\frac{\pi}4\cop_3\cop_4\right)}
\exp{\left(-\frac{\pi}4\cop_5\cop_6\right)}.
\end{eqnarray}
The last two exponents in Eq.~(\ref{cZ}) are braid gates, so
the controlled $\sm^z$ gate is equivalent to $O3$ (we disregard
the overall phase).
One remains to prove the lemma.

{\bf Proof of Lemma~\ref{lemma:non-linear}:}

\noindent
{\it $O3$ can simulate $O1$}:
Using solely braid gates one can prepare a state
with a stabilizer group
\[
S=(-i\cop_1\cop_7, \; -i\cop_2\cop_8, \; -i\cop_3\cop_5, \;
-i\cop_4\cop_6)
\]
(all eigenvalues are $+1$).
Let this state be acted upon by an operator
\[
U\equiv \exp{(i\frac{\pi}4 \cop_1\cop_2\cop_3\cop_6)}.
\]
The conjugated action of $U$ on the Majorana operators is as follows:
\[
U\cdot U^\dag \, : \, \left\{ \ba{rclrcl}
\cop_1 &\to & -i\cop_2\cop_3\cop_6, &
\cop_5 &\to & \cop_5, \\
\cop_2 &\to & i\cop_1\cop_3\cop_6, &
\cop_6 &\to & i\cop_1\cop_2\cop_3, \\
\cop_3 &\to & -i\cop_1\cop_2\cop_6, &
\cop_7 &\to & \cop_7, \\
\cop_4 &\to & \cop_4, &
\cop_8 &\to & \cop_8. \ea \right.
\]
Accordingly, the stabilizer group $S$ is mapped into
\[
S'=(-\cop_2\cop_3\cop_6\cop_7, \cop_1\cop_3\cop_6\cop_8,
-\cop_1\cop_2\cop_5\cop_6, -\cop_1\cop_2\cop_3\cop_4).
\]
It coincides with the stabilizer group $(S_1,S_2,S_3)$
of the state $|a_8\ra$, see Eq.~(\ref{Ss}). Therefore
the state stabilized by $S'$ coincides with $|a_8\ra$
up to an overall phase.

\begin{center}
\begin{figure}
\includegraphics[scale=0.3]{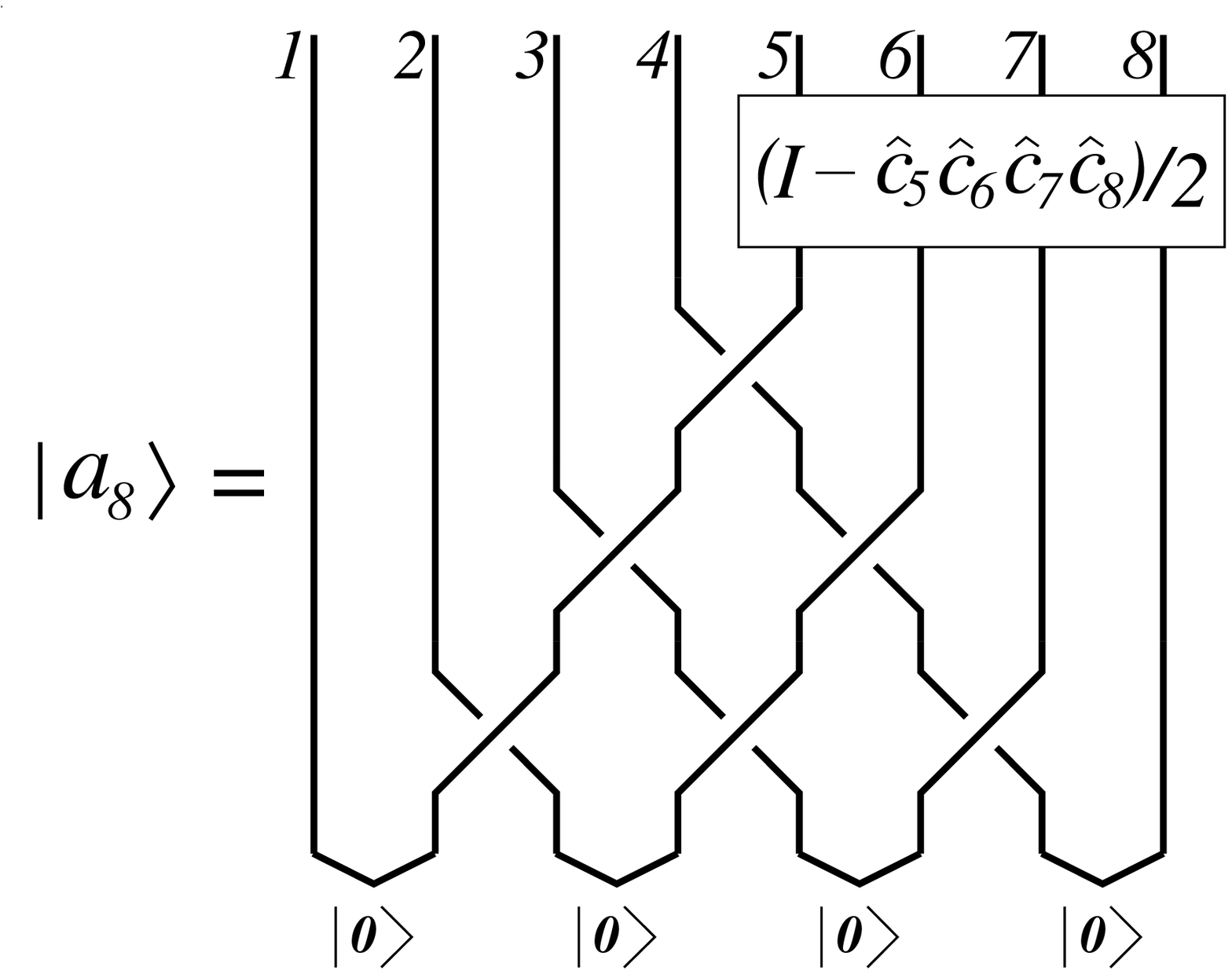}
\caption{Preparation of $|a_8\ra$ via eigenvalue measurement of $\cop_5\cop_6\cop_7\cop_8$.}
\label{fig:psi8}
\end{figure}
\end{center}

\noindent
{\it $O2$ can simulate $O1$}:
(This part is not necessary for the proof,
but we shall use this result later.)
Using solely braid gates one can prepare a state
with a stabilizer group
\[
S=(-i\cop_1\cop_5, \; i\cop_2\cop_6, \; -i\cop_3\cop_7, \;
i\cop_4\cop_8).
\]
Let us measure an eigenvalue of $-\cop_5\cop_6\cop_7\cop_8$
on this state. To find the final stabilizer group, choose
generators of $S$ as
\[
S=(-\cop_1\cop_2\cop_5\cop_6,  -\cop_2\cop_3\cop_6\cop_7,
-\cop_3\cop_4\cop_7\cop_8, i\cop_4\cop_8).
\]
After the measurement the last generator is replaced
by $-\cop_5\cop_6\cop_7\cop_8$ (may be with the opposite sign),
which is equivalent to a generator $-\cop_1\cop_2\cop_3\cop_4$.
The resulting stabilizer group coincides with the one of
$|a_8\ra$. The preparation of $|a_8\ra$ by
measuring $\cop_5\cop_6\cop_7\cop_8$ is
illustrated by Fig.~\ref{fig:psi8}.

\begin{center}
\begin{figure}
\includegraphics[scale=0.35]{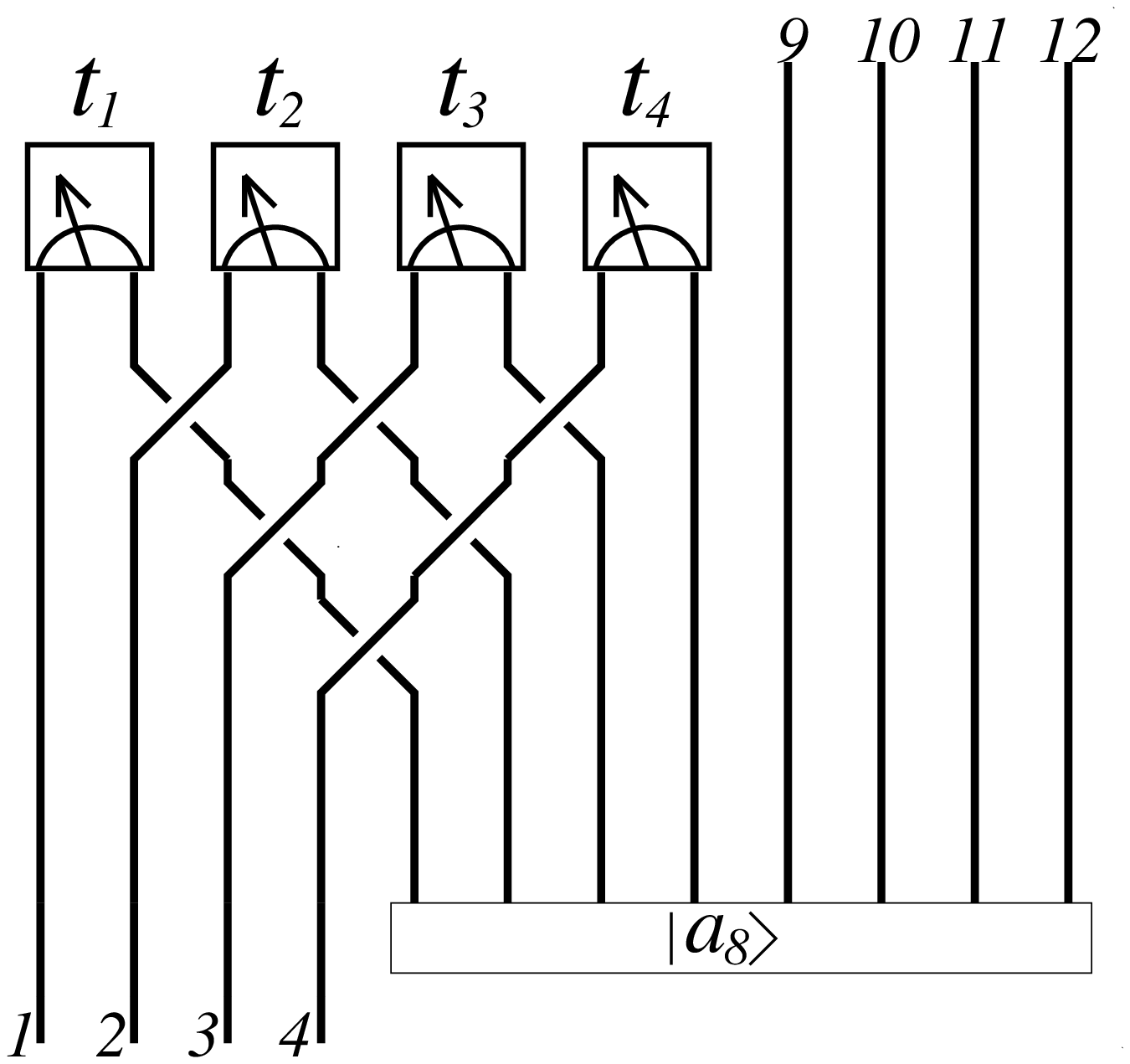}
\caption{Implementation of a non-destructive measurement of
$\cop_1\cop_2\cop_3\cop_4$ that consumes
 one copy of $|a_8\ra$.}
\label{fig:cccc}
\end{figure}
\end{center}

\noindent
{\it $O1$ can simulate $O2$}:
Assume that one copy of $|a_8\ra$ is available.
A sequence of braidings and fusions
that allows one to measure an eigenvalue of
$\cop_1\cop_2\cop_3\cop_4$ is shown on
Fig.~\ref{fig:cccc}. The particles labeled by $1,2,3,4$
on the figure are prepared in an arbitrary initial
state $|\psi_{in}\ra$.
The state $|a_8\ra$ is described by generators $\cop_5,\ldots,\cop_{12}$.
Accordingly, the circuit shown on Fig.~\ref{fig:cccc}
is applied to a state $|\psi_{in}\otimes a_8\ra$.
The particles are reshuffled by a braid operator
and then
observables $T_1=-i\cop_1\cop_2$, $T_2=-i\cop_3\cop_4$,
$T_3=-i\cop_5\cop_6$, and $T_4=-i\cop_7\cop_8$ are measured.
The final state $|\psi_f\ra$ is read out from the particles
$9,10,11,12$.

Let $(-1)^{t_j}$ be the measured eigenvalue of $T_j$.
We shall consider in details
only the case $t_1\oplus t_2\oplus t_3\oplus t_4=0$.
Let $|\Phi_{\tt}\ra$ be the  final state
corresponding to outcomes $\tt=(t_1,t_2,t_3,t_4)$.
Using the stabilizer description of $|a_8\ra$,
see Eq.~(\ref{Ss}), one can easily check that
\[
|\Phi_{\tt}\ra=(\cop_2^{t_1}\, \cop_4^{t_2}\,
\cop_6^{t_3}\, \cop_8^{t_4}) (\cop_9^{t_1}\,
\cop_{10}^{t_2}\, \cop_{11}^{t_3}\, \cop_{12}^{t_4})
|\Phi_{\bf 0}\ra
\]
whenever $t_1\oplus t_2\oplus t_3\oplus t_4=0$.
Let us apply additional braid gates $\cop_2\cop_9$,
$\cop_4\cop_{10}$, $\cop_6\cop_{11}$, and $\cop_8\cop_{12}$
controlled by classical bits $t_1,t_2,t_3$, and $t_4$
respectively (these gates are not shown on Fig.~\ref{fig:cccc}
to avoid clutter).
They map $|\Phi_{\tt}\ra$ into $|\Phi_{\bf 0}\ra$,
so it suffices to analyze the case $t_j=0$.
Obviously, $|\Phi_{\bf 0}\ra$ has a product structure
$|\Phi_{\bf 0}\ra=|0,0,0,0\ra\otimes |\psi_{fin}\ra$.

\begin{center}
\begin{figure}
\includegraphics[scale=0.3]{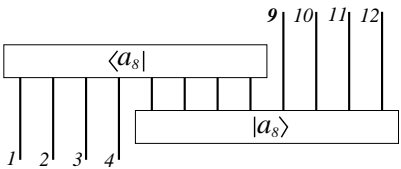}
\caption{Teleportation.}
\label{fig:teleportation}
\end{figure}
\end{center}

One can easily notice that
the left-upper part
of Fig.~\ref{fig:cccc} with $t_j=0$
is almost identical to
the preparation procedure for $\la a_8|$, see
Fig.~\ref{fig:psi8}.
The only missing element is
a projector $(1/2)(I-\cop_5\cop_6\cop_7 \cop_8)$.
However,
one can safely add the missing projector because
it stabilizes the state $|a_8\ra$.
Therefore, for the outcomes $t_j=0$ the protocol
shown on
Fig.~\ref{fig:cccc} coincides with the one shown on
Fig.~\ref{fig:teleportation}.
Taking into account that $|a_8\ra$ is the encoded
EPR state, $|a_8\ra=2^{-1/2}(|\bar{0},\bar{0}\ra+|\bar{1},\bar{1}\ra)$,
the protocol on FIG.~\ref{fig:teleportation}
is just a projection of $|\psi_{in}\ra$ onto the code
subspace followed by
teleportation of the encoded qubit
from the particles $1,2,3,4$ to the particles $9,10,11,12$.
Accordingly, $|\psi_f\ra=(1/2)(I-\cop_1\cop_2\cop_3
\cop_4)|\psi_{in}\ra$, up to an overall
normalization constant.
Using similar arguments one can check that
$|\psi_f\ra=(1/2)(I+\cop_1\cop_2\cop_3
\cop_4)|\psi_{in}\ra$ whenever
$t_1\oplus t_2\oplus t_3\oplus t_4=1$.

\noindent
{\it $O2$ can simulate $O3$}:
(This result has been already proved in~\cite{BK00f}.)
Suppose we want to implement an operator $\exp{(i\frac{\pi}{4}
\cop_1\cop_2\cop_3\cop_4)}$.
Let us prepare an ancillary pair of particles
$5,6$ in the state $|0\ra$. Accordingly,
any input state $|\psi\ra$ of the system satisfies
\be\label{initcond}
(\cop_5+i\cop_6)\, |\psi\ra=0.
\ee
Let us measure an eigenvalue of
$\cop_1\cop_2\cop_4\cop_5$.
Depending upon the outcome,
the initial state  $|\psi\ra$ gets multiplied by a projector
$\Pi^{(4)}_{\pm}=(1/2)(I\pm \cop_1\cop_2\cop_4\cop_5)$
(with a proper normalizing coefficient).
Next we measure an
eigenvalue of $-i\cop_3\cop_5$. The  eigenvalues $\pm 1$ correspond to
projectors $\Pi^{(2)}_{\pm}=(1/2)(1\mp i\cop_3\cop_5)$.
We claim that after some correction
depending on the measurements outcomes, the protocol effectively executes the
operator  $\exp{(i\frac{\pi}{4} \cop_1\cop_2\cop_3\cop_4)}$
while leaving the ancillary pair of particles intact.
The correction step requires only braid gates.
Indeed, one can use the following identities:
\be
\ba{r}
\multicolumn{1}{l}{
\exp{\left(i\frac{\pi}{4}\cop_1\cop_2\cop_3\cop_4\right)}
\, |\psi\ra= }\\
= 2\exp\left(\frac{\pi}{4}\cop_3\cop_6\right)
\Pi^{(2)}_{+} \Pi^{(4)}_{+}
\, |\psi\ra\\
=2i\exp{\left( \frac{\pi}2 \cop_1\cop_2\right)}
\exp{\left( \frac{\pi}2 \cop_3\cop_4\right)}
\exp\left(\frac{\pi}{4}\cop_3\cop_6\right)
\Pi^{(2)}_{-}\Pi^{(4)}_{-}
\, |\psi\ra \\
=\ 2i\exp{\left( \frac{\pi}2 \cop_1\cop_2\right)}
\exp{\left( \frac{\pi}2 \cop_3\cop_4\right)}
\exp\left(-\frac{\pi}{4}\cop_3\cop_6\right)
\Pi^{(2)}_{-}\Pi^{(4)}_{+}
\, |\psi\ra \\
=2\exp\left(-\frac{\pi}{4}\cop_3\cop_6\right)
\Pi^{(2)}_{-}\Pi^{(4)}_{-}
\, |\psi\ra\\
\end{array}
\end{equation}
(we have used Eq.~(\ref{initcond})).
In each of the four cases one can apply a suitable correction
operator $U_{yz}$
(for example,
$U_{++}=\exp(\frac{\pi}{4}\cop_3\cop_6)$
if the outcomes were $++$, etc.) so that
\[
\exp{\left(i\frac{\pi}{4}
\cop_1\cop_2\cop_3\cop_4\right)} \, |\psi\ra
 \,=\,
2\,U_{yz}\Pi^{(2)}_{y}\Pi^{(4)}_{z}\,|\psi\ra.
\]
Each of the four outcome combinations occurs with probability
$1/4$. The final state is always the desired one,
i.e., $\exp{(i\frac{\pi}{4}
\cop_1\cop_2\cop_3\cop_4)} \, |\psi\ra$.

\section{Universal quantum computation}\label{sec:UQC}

The protocols described in Section~\ref{sec:Clifford} allow one
to execute any Clifford group gates on the logical qubits.
In addition to that, one can measure logical qubits in the standard
basis and prepare fresh logical qubits in the state $|\bar{0}\ra$.
Let us refer to this set of operations as {\it Clifford operations}.
As we have shown, Clifford operations can be implemented with an arbitrarily
small error rate and the overhead is polylogarithmic. To simplify the
discussion, we shall firstly set the error rate of
Clifford operations to zero and then
address the precision and overhead issues
separately.

Below we will show how to execute the $\pi/8$-rotation
\be\label{pi/4}
\Lambda(\phase)=\left( \ba{cc} 1 & 0 \\ 0 & \phase \\ \ea \right)
\ee
on the logical qubit.
It is well known that the $\pi/8$-rotation together with Clifford
operations constitute a universal set of gates.

Although the $\pi/8$-rotation cannot be implemented by
Clifford operations only,
we can follow the same strategy as in Section~\ref{sec:a8},
namely,  try to use very noisy non-topological operations to prepare
a state $\rho$ that approximates some
logical target state $|a\ra$, improve an accuracy of the
approximation by running a purification protocol
(that now can use any Clifford operations), and then
convert $|a\ra$ into
the gate $\Lambda(\phase)$.

It is not apriori clear what ancillary state $|a\ra$ leads to the
most efficient implementation of $\Lambda(\phase)$.
We shall argue that a good choice of $|a\ra$ is a state
\[
|a_4\ra=\frac1{\sqrt{2}} ( |\bar{0}\ra + \phase \, |\bar{1}\ra)=
 \Lambda(\phase)\, |\bar{+}\ra,
\]
where $|\bar{+}\ra=2^{-1/2}(|\bar{0}\ra+|\bar{1}\ra)$.
The state  $|a_4\ra$ is composed from four $\sigma$-particles.
A purification protocol for $|a_4\ra$ with
a high threshold error rate and polylogarithmic overhead which uses only
Clifford operation has been put forward in~\cite{BK04a}
under the name ``magic states distillation''.
For the sake of completeness we briefly describe it below.
Then we assess an efficiency of the whole simulation
scheme.

In the rest of this section
a word `qubit' refers to a logical qubit encoded by
four $\sigma$-particles as explained in Section~\ref{sec:TQC}.
By abuse of notations we shall abbreviate $|\bar{0}\ra$ to
$|0\ra$ and $|\bar{1}\ra$ to $|1\ra$.
Accordingly, $|a_4\ra$
will be regarded as a one-qubit state.

\subsection{Converting $|a_4\ra$ into a non-Clifford gate}
We start from explaining how to execute the gate
$\Lambda(\phase)$ using Clifford operations
and one copy of $|a_4\ra$, see~\cite{BK04a}.
Let $|\psi\ra=a|0\ra+b|1\ra$ be an unknown
state (the coefficients $a$ and $b$ may
actually be quantum states as well).
Suppose we want to apply the gate $\Lambda(\phase)$ to $|\psi\ra$.
Let us start from a two-qubit state $|\Psi_0\ra=|\psi\otimes
a_4\ra$ and measure an eigenvalue of observable
$T_1=\sigma^z\otimes \sigma^z$ (recall that any multi-qubit
Pauli operator can be converted by Clifford gates into
one-qubit operator $\sm^z$, which is an admissible observable
in the TQC model).
The outcomes $\pm 1$ of the measurement appear
with the probability $1/2$ each, yielding the  final states
\begin{eqnarray}
|\Psi_1^{+}\ra&=&a |0,0\ra + b \phase |1,1\ra, \nn \\
|\Psi_1^{-}\ra&=& a\phase |0,1\ra + b  |1,0\ra.
\end{eqnarray}
Applying the controlled $\sm^x$ operator $\Lambda(\sm^x)$
with the first qubit as a control one, we get
\begin{eqnarray}
|\Psi_2^{+}\ra&=&\Lambda(\sm^x)|\Psi_1^{+}\ra
=(a |0\ra + b \phase |1\ra)\otimes |0\ra, \nn \\
|\Psi_2^{-}\ra&=&\Lambda(\sm^x)|\Psi_1^{-}\ra
=(a \phase |0\ra + b |1\ra)\otimes |1\ra. \nn
\end{eqnarray}
Now let us measure the second qubit in the
$\{|0\ra,|1\ra\}$ basis. If the outcome is $|1\ra$,
apply additional Clifford gate
$K=|0\ra\la 0| + i \, |1\ra\la 1|$
to the first qubit (as was mentioned in
Section~\ref{sec:TQC},  $K$ is a braid gate).
In both cases we end up with  the final
state $a|0\ra + b\phase |1\ra$. Thus the
input state $|\psi\ra$ has been acted upon by
$\Lambda(\phase)$.

\subsection{Purification of $|a_4\ra$}

Here we outline the magic states distillation method,
see the original paper~\cite{BK04a} for details.
A noisy $|a_4\ra$ state will be described by a
one-qubit density matrix $\rho$. A quality of
$\rho$ is characterized by a parameter
\[
\ep=1-\la a_4|\rho|a_4\ra
\]
which will be referred to as an {\it error rate}.
The purification protocol exploits some nice properties
of the CSS second-order punctured Reed-Muller quantum
code. It encodes one qubit into $15$ qubits
and has the minimal distance $3$. Let $\Pi$ be a
projector on the code subspace of the Reed-Muller code.
Consider a state
\[
\rho_{out}=Z^{-1}\, \Pi\, \rho^{\otimes 15} \, \Pi,
\quad Z\equiv \trace{(\Pi\, \rho^{\otimes 15})}.
\]
Although $\rho_{out}$ is a $15$-qubit state, it can
be regarded as a one-qubit state encoded by the
Reed-Muller code.
It turns out that an error rate $\ep_{out}$ of
the state $\rho_{out}$ is cubically
suppressed as compared to the error rate of $\rho$,
\[
\ep_{out}= 35\, \ep^3 + O(\ep^4).
\]
The properties of the Reed-Muller quantum code that
are responsible for this effect are (i) The minimum
Hamming weight of $\sm^z$-type errors that are not
detected by the code is $3$; (ii) The code has non-Clifford
automorphisms: an operator $\Lambda(\phase)^{\otimes 15}$
commutes with $\Pi$ and its action on the encoded qubit
coincides with $\Lambda(\phase)$.

This observation provides a natural mean of
purifying $\rho$.
Namely, one takes $15$ copies of $\rho$ and
measures eigenvalues of $14$ stabilizer operators
for the Reed-Muller code. All stabilizers
are the Pauli operators, so these measurements
require only Clifford gates and
admissible TQC measurements.
The final state
is accepted iff one observes the trivial syndrome
(eigenvalue of all stabilizer operators is $+1$).
After that one applies a decoding transformation
(a certain Clifford group operator) that
maps $\rho_{out}$ into a one-qubit state.
The threshold value of  $\ep$ is determined by
an equation $\ep_{out}(\ep)=\ep$. Denote the threshold
by $\delta_4$. Its numerical value is
\[
\delta_4\approx 0.141.
\]
If $\ep<\delta_4$, the output state $\rho_{out}$ is more clean than
the input one, i.e., $\ep_{out}(\ep)<\ep$.

Let $p_s$ be
the probability for this algorithm to succeed, i.e.,
the probability to observe the trivial syndrome.
In the limit $\ep\to 0$ one has $p_s\approx 2^{-10}$.
Moreover, by introducing an additional
``error correction'' step into the algorithm one
can accept a larger set of measured syndromes (syndromes
for which only all $\sm^x$-type stabilizers have eigenvalue $+1$).
The error correction step
enhances the success probability to
$p_s\approx 1$ (in the limit $\ep\to 0$).

The initial supply of states $\rho$ with an error
rate $\ep_0\equiv \ep$ will be called level-$0$
ancillas.
Accordingly, level-$k$ ancillas are obtained from
the level-$0$ ancillas by iterating the elementary
purification procedure $k$ times. Let $\ep_k$ and
$n_k$ be an error rate and the total number of
level-$k$ ancillas. The numbers $\ep_{k+1}$, $n_{k+1}$
and $\ep_k$, $n_k$ are related by recursive flow equations
\be\label{a4:flow}
n_{k+1}\approx \frac{n_k}{15}, \quad
\ep_{k+1}\approx 35\, \ep_k^3
\ee
(we are interested in the asymptotic regime $\ep\ll 1$).
Accordingly,
if one needs to prepare one copy of $|a_4\ra$ with
an error rate $\ep'$, one needs to have a supply of
\be\label{a4:efficiency}
n_0\sim |\log(\ep')|^\gamma,\quad
\gamma = \log_3{15} \approx 2.5
\ee
level-$0$ ancillas with an error rate below the threshold,
$\ep_0<\delta_4$. The operational cost of the purification,
i.e., the total number of Clifford gates
and standard measurements needed to achieve an error rate
$\ep'$ has the same scaling as $n_0$.

\subsection{Efficiency analysis}
Suppose our goal is to simulate a quantum circuit with
$N$ one-qubit and two-qubit gates operating on $n$ qubits.
We assume that the following gate set is used:
\be\label{gate_set}
\frac1{\sqrt{2}} \left(\ba{cc} 1 & i \\ i & 1 \\ \ea\right),
\quad \Lambda(\phase), \quad \Lambda(\sigma^z).
\ee
The simulation must be able to reproduce the output of the circuit
with a constant error probability.
Accordingly, the non-topological gates
$\Lambda(\phase)$ and $\Lambda(\sigma^z)$ have to be
simulated with an error probability $\delta\sim N^{-1}$.
As we have learned in Section~\ref{sec:a8},
preparation of $|a_8\ra$  with an
accuracy $\delta$
requires about $(\log{(\delta^{-1})})^3$
raw ancillas $\rho_8$ and about the same number of
TQC operations. According to Section~\ref{sec:Clifford}
one copy of $|a_8\ra$ can be traded for
a gate $\Lambda(\sigma^z)$ implemented with about the same precision.
Thus each $\Lambda(\sigma^z)$ gate
`costs' $O((\log{(N)})^3)$ TQC operations and raw ancilla
preparations.

Simulation of the gate $\Lambda(\phase)$ deserves more
careful analysis. Consider
one round of $a_4$-purification at the level $k$.
It takes as input $15$ copies of level-$k$ ancillas $|a_4\ra$
with an error rate $\ep_k$
and  outputs one copy of $|a_4\ra$ with an error rate $\ep_{k+1}$
(sometimes it outputs nothing because we use postselection).
An implementation of this $a_4$-purification round requires $O(1)$ gates $\Lambda(\sigma^z)$.
To simulate each of these gates the $a_8$-purification protocol has to be
invoked. Obviously, at this point it does not make sense to purify $|a_8\ra$
ancillas all way down to the error rate $\delta\sim N^{-1}$.
Instead, the error rate $O(\ep_k^3)$ is sufficient, since it
still gives the flow equation $\ep_{k+1}=C\ep_k^3$ for $a_4$-purification
with some  constant $C$.
Comparing Eq.~(\ref{a8:efficiency}) and Eq.~(\ref{a4:efficiency})
one can see that for a fixed error rate
the simulation of $\Lambda(\sigma^z)$ is more
demanding in terms of resources than the simulation
of $\Lambda(\phase)$. Therefore, we can try to use the
above observation to improve the efficiency of the whole
simulation scheme.

Indeed, purification of one copy of $|a_8\ra$ with the final error rate
$O(\ep_{k}^3)$
requires
$m_k \sim (\log{(\ep_{k})})^3$
elementary operations.
From Eq.~(\ref{a4:flow}) one gets
$\ep_{k}\sim \exp{(-c3^k)}$, where $c>0$ is a constant.
Therefore, $m_k\sim 3^{3k}$.
The total number of $a_4$-purification
rounds on the level $k$ is $g_{k}\approx n_{k+1}\approx n_0 15^{-k-1}$,
where $n_0$ is the number of level-$0$ ancillas $|a_4\ra$.
Thus the total number of elementary operation
needed to generate all level-$(k+1)$ ancillas $|a_4\ra$
is $M_k=m_k g_k \sim n_0 15^{-k} 3^{3k}$.
Clearly, $M_k$ grows exponentially with $k$,
so almost all resources needed to purify $|a_4\ra$
are spent at the highest level of $a_4$-purification.
Accordingly,  the total number of elementary operations
needed to purify one copy of $|a_4\ra$ with the final error rate
$\delta\sim N^{-1}$ is
\[
M_{tot}=\sum_{k=1}^d M_k\approx M_d
\approx 3^{3d}, \quad d\approx \log_{15}{(n_0)}.
\]
From Eq.~(\ref{a4:efficiency}) with $\delta\sim N^{-1}$ one
gets $n_0\approx (\log{(N)})^\gamma$. Therefore
\[
M_{tot}\sim (\log{(N)})^3.
\]
We conclude that any gate in the universal gate set Eq.~(\ref{gate_set})
`costs' $O((\log{(N)})^3)$ elementary operations.

\section{Implementation of non-topological operations}
\label{sec:nto}

This part of the paper is rather speculative, since we know almost nothing
about non-topological properties of anyons, such as
effects of a finite separation between particles, non-adiabaticity of
the anyonic transport, interaction between an anyon and
a control device, e.t.c..

Recall that we need non-topological operations to prepare
(may be very noisy)
ancillary states $|a_4\ra$ and $|a_8\ra$ composed from four and eight
$\sigma$-particles respectively.
We shall argue below that a good strategy is
 to use direct short-range interaction between anyons. One can
expect that the amplitude of this interaction decays as
$\exp{(-l/l_H)}$, where $l$ is a separation between the particles
and $l_H$ is the magnetic length (for experiments with
AlGaAs/GaAs heterostructures the magnetic field corresponding to
$\nu=5/2$ is $B\approx 5$T, so that $l_H=(\hbar\,  c/e B)^{1/2} \approx
10^{-6}$cm).

{\it Remark:} Note that any state in the orbit of $|a_4\ra$ or $|a_8\ra$
under the action of braid gates is equally acceptable
as the states
$|a_4\ra$ or $|a_8\ra$ themselves.
As the number of particles
increases, the size of the orbit grows, and thus the set of
acceptable states becomes larger.
For example, the orbit of $|a_4\ra$
consists of $12$ states (we disregard the overall phase),
while
the orbit of $|a_8\ra$
consists of $240$ states~\footnote{This counting goes along the
following lines: (1) The set of four-qubit stabilizer states with a
fixed parity consists of two non-overlapping subsets: the orbit of
$|a_8\ra$ and the subset of `paired' states whose stabilizer group
can be represented as in Eq.~(\ref{paired_stabilizer_group});
(2) The number of four-qubit stabilizer states with a fixed parity
(say $+1$) is equal to the total number of three-qubit stabilizer
states; (3) There are totally $1080$ three-qubit stabilizer states;
(4) There are totally $2^3 \, 8!!=840$ `paired' four-qubit stabilizer
states with a fixed parity (say $+1$). Therefore the number of states
in the orbit of $|a_8\ra$ is $1080-840=240$.}.
It suffices to prepare any of
these states (we have to know which)
with a fidelity above the threshold one.

\subsection{How to prepare  $|a_4\ra$}
\begin{center}
\begin{figure}
\includegraphics[scale=0.32]{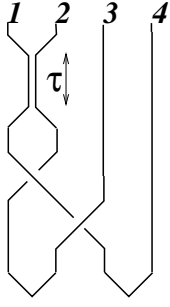}
\caption{A preparation of $|a_4\ra$ based on the  short-range
two-particle interaction.} \label{fig:shortrange1}
\end{figure}
\end{center}

Let us start from preparation of $|a_4\ra$ since it is much easier.
The preparation process is illustrated on Fig.~\ref{fig:shortrange1}.
One starts from the vacuum state, creates two pairs of
$\sigma$-particles and then brings two particles, one
from each pair, sufficiently close to each other. After that one waits
for a time $\tau$ and finally returns the particles to their original
positions.

Taking into account that  the short-range interaction is a local
operator, we infer that the total charge of the four particles
and the total charge of the particles $1$, $2$ must be preserved.
Therefore, the interaction can be described by a Hamiltonian
\[
H_{int}=-i\cop_1\cop_2\otimes X + I\otimes Y,
\]
where $X$ and $Y$ are some operators acting on the environment.

A purpose of the two braid operations preceding the interaction on Fig.~\ref{fig:shortrange1}
is to create a state
\[
|\phi\ra=B_{1,2}^\dag \, B_{2,3} \, |0,0\ra = 2^{-1/2}\, \left(
 |0,0\ra + |1,1\ra \right) \in \calF_4.
\]
Using the qubit representation of Section~\ref{sec:TQC}
one gets $|\phi\ra=2^{-1/2}\, (|\bar{0}\ra + |\bar{1}\ra)$.

Free evolution under $H_{int}$ for the time $\tau$ maps $|\phi\ra$ into
a state
\[
\frac1{\sqrt{2}} \left( |\bar{0}\ra \otimes e^{i(X+Y)\tau}\,
|\Psi_E\ra + |\bar{1}\ra \otimes e^{i(-X+Y)\tau}\, |\Psi_E\ra\right).
\]
Here $|\Psi_E\ra$ is the initial state of the environment (one can always
assume that it is pure).
Tracing out the environment we end up with a mixed state
\[
\rho=\frac12 \left( \ba{cc} 1 & r \\ r^* & 1 \\ \ea \right),
\;
r=\la \Psi_E |e^{i(X+Y)\tau}\, e^{i(X-Y)\tau}|\Psi_E\ra.
\]
The case $\rho=|a_4\ra\la a_4|$ corresponds to $r=e^{i\pi/4}$.
By varying the interaction time $\tau$ we can try to fulfill the
threshold condition $\la a_4|\rho|a_4\ra > 1-\delta_4\approx 0.86$.
This may or may not be possible, depending upon particular form of
$X$, $Y$, and $|\Psi_E\ra$. For example, if $X$ is proportional
to the identity operator, $X=g\, I$, one gets
\[
r=e^{2ig\tau}.
\]
Tuning $\tau$ such that $g\tau=\pm \pi/8$ we can prepare
the desired state $|a_4\ra$ (or a state that can be converted to
$|a_4\ra$ by a braid gate).

\subsection{How to prepare $|a_8\ra$}

Preparation of $|a_8\ra$ based on the direct short-range interaction
between anyons is more tricky because one has to cancel unwanted
interactions. For example, if $\sigma$-particles $1,2,3,4$ are
sufficiently close to each other, the interaction Hamiltonian looks as
\[
H=-i \sum_{j,k} \cop_j \cop_k \otimes X_{jk} -
 \cop_1\cop_2\cop_3\cop_4 \otimes X + I\otimes Y,
\]
where $X_{jk}$, $X$, and $Y$ are some operators acting on the
environment.
Recall that $|a_8\ra$ can be prepared by TQC operations and a
non-linear gate $W=\exp{(i\pi/4 \, \cop_1\cop_2\cop_3\cop_4)}$, see
the first part of the proof of Lemma~\ref{lemma:non-linear}.
Free evolution under the Hamiltonian $H$ might be used to implement
$W$, provided that one can ``turn off'' the quadratic interactions
$\cop_j\cop_k\otimes X_{jk}$. In principle, it can be done using
a technique analogous to decoupling and refocusing in the Nuclear Magnetic
Resonance.
Indeed, denote $F=\cop_1\cop_2$, $G=\cop_1\cop_3$ and consider a
Hamiltonian
\[
H' = \frac14 (H + F\, H \, F^\dag + G \, H \, G^\dag + (FG)\, H \,
(FG)^\dag).
\]
One can easily check that
\[
H'= - \cop_1\cop_2\cop_3\cop_4 \otimes X + I \otimes Y.
\]
Now let $U_{\tau}$ and $U_{\tau}'$ be unitary operators
describing evolution under the Hamiltonians $H$ and $H'$ respectively
for a time $\tau$. If $\tau$ is sufficiently small, one gets
from the Trotter expansion
\[
U_{\tau}' \approx U_{\frac{\tau}4} \cdot \left( F U_{\frac{\tau}4}
F^\dag\right)\cdot
\left(  G U_{\frac{\tau}4} G^\dag\right)
\cdot
\left( FG U_{\frac{\tau}4} G^\dag F^\dag \right).
\]
Therefore one could try to simulate $U_{\tau}'$ by $U_{\tau/4}$
and ``control pulses'' $F$ and $G$. Obviously, $F$ and $G$ can
be implemented by braid gates (for example, $F$ corresponds to
winding particle $1$ around the particle $2$). However,
before applying any of these braid gates one has to return the
particle $1,2,3,4$ into original well-separated positions.
After that one can compose the evolutions $U_{\tau}'$ to simulate
any desired interaction time.

The preparation of $|a_8\ra$ based on the refocusing
may fail to provide the necessary precision $\delta_8$
because it involves too many noisy operations. So it is
more fair to say that an additional non-topological operation
is needed.

According to Lemma~\ref{lemma:non-linear}, the state
$|a_8\ra$ can also be prepared by TQC operations and
a non-destructive measurement of an observable
$\cop_1 \cop_2 \cop_3 \cop_4$.
In other words, we have to measure the total topological
charge ($\one$ or $\ppsi$) of four $\sigma$-particles
without destroying their pairwise correlations.
It is very likely that such a measurement can be
implemented using an interferometric device proposed
recently by Bonderson, Kitaev, and Shtengel~\cite{Bon05a}
(see also~\cite{FNTW98a})
to test topological properties of $\sigma$-particles.

\begin{center}
\begin{figure}[h]
\includegraphics[scale=0.3]{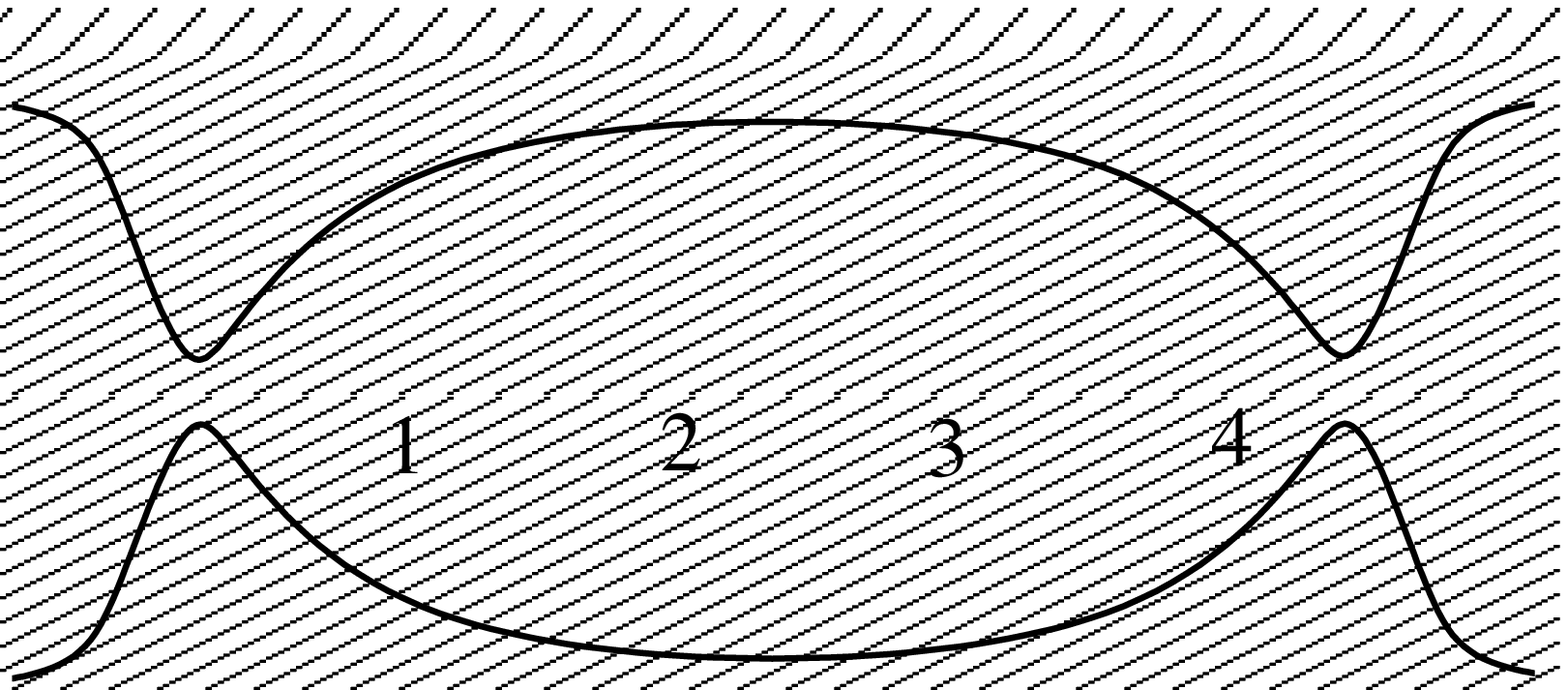}
\caption{A two-point contact interferometer.
Four $\sigma$-particles are trapped at antidots inside the
interferometer loop. Electric current propagates
along the top and bottom edges of the FQH electron gas
(the unshaded region). Tunneling of $\sigma$-particles
occurs at the constrictions.}\label{fig:Hall_bar}
\end{figure}
\end{center}

The device is based on the Hall bar geometry, see
Fig.~\ref{fig:Hall_bar}, so that the transport of electric charge
is governed by edge currents on the top and
bottom edges of the bar. Electrical gates are used to create
two constrictions in the region occupied by the FQH electron gas
(the unshaded region on Fig.~\ref{fig:Hall_bar}), so that
$\sigma$-particles can tunnel between the top and bottom edges
through the electron gas at either constriction.
The parameters of the device are tuned to allow
quantum  interference between the two tunneling paths.
The total tunneling current is measured through
the longitudinal resistance $R_{xx}$.
Ideally, such a measurement projects the initial state
onto an eigenvector of the tunneling current operator.

\begin{center}
\begin{figure}[h]
\includegraphics[scale=0.25]{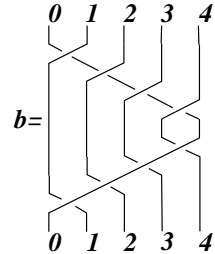}
\caption{A braid $b$ describing interference of the two tunneling
paths
in the two-point contact interferometer.}\label{fig:Hall_bar_1}
\end{figure}
\end{center}

Suppose that four antidots are created inside the interferometer loop,
and exactly one $\sigma$-particle is trapped at each antidot.
Let us label the trapped $\sigma$-particles by $1,2,3,4$ and
the tunneling $\sigma$-particle by $0$.
The difference between the two tunneling paths
corresponds to a braid $b$ in which the tunneling particle $0$
winds around the trapped particles $1,2,3,4$, see Fig.~\ref{fig:Hall_bar_1}.
Using  the braid group representation described in Section~\ref{sec:Ising}
one can easily find that the action of $b$ is
$\varphi(b)=+\cop_1\cop_2\cop_3\cop_4$.
Thus the  longitudinal resistance measurement projects
the initial state of the particles $1,2,3,4$ onto
an eigenvector of $\cop_1\cop_2\cop_3\cop_4$.
Combining the interferometric experiment with the standard
TQC measurements one can calibrate the device to
infer an eigenvalue of $\cop_1\cop_2\cop_3\cop_4$
from the measurement outcome.

\begin{acknowledgements}

Discussions with Alexei Kitaev and Robert Raussendorf are gratefully acknowledged.
The author would like to thank David DiVincenzo for
a careful reading of this paper, and
Andrei Soklakov for useful
comments concerning the magic states distillation protocol.
This work was supported by the National Science Foundation under
grant number EIA-0086038.

\end{acknowledgements}

\end{document}